\pdfoutput=1

\documentclass[letterpaper,twocolumn,10pt]{article}
\usepackage{preprint}

\usepackage{amsmath}

\usepackage{color}
\usepackage{graphicx}
\usepackage[labelformat=simple]{subcaption}
\usepackage{xspace}
\usepackage{enumitem}
\setlist[1]{leftmargin=1.5em}
\setlist[2]{leftmargin=1.5em}

\usepackage{pifont} %

\usepackage{bm}
\usepackage[spaceRequire=false]{algpseudocodex}     %
\algrenewcommand\algorithmicrequire{\textbf{Input}} %
\algnewcommand{\algorithmicyield}{\textbf{yield}\xspace}
\algnewcommand{\Yield}{\algorithmicyield}

\usepackage{algorithm}

\usepackage{xcolor}
    \definecolor{lightgray}{gray}{0.9} %
    \definecolor{mediumgray}{gray}{0.8} %
    \definecolor{darkgray}{gray}{0.4} %

\usepackage{array}
\usepackage{tabularx}
    \newcolumntype{L}[1]{>{\raggedright\arraybackslash}p{#1}}
    \newcolumntype{R}{>{\raggedleft\arraybackslash}X}
    \newcolumntype{C}{>{\centering\arraybackslash}X}

\usepackage{multirow} %
\usepackage{makecell} %
\usepackage{diagbox} %
\usepackage{booktabs} %
\usepackage{colortbl} %

\usepackage{ulem}
\usepackage{hyperref}
\usepackage{xurl} %

\usepackage{tikz}
\usetikzlibrary{positioning, arrows.meta, fit, calc, shapes.geometric, backgrounds, shadows.blur}
\usepackage{pgfplots}
\pgfplotsset{compat=1.18}
\tikzset{
  mj/box/.style={
    draw=black!55, rectangle, rounded corners=2pt, align=center,
    minimum height=6mm, inner sep=3pt, fill=white, font=\scriptsize
  },
  mj/agent/.style={mj/box, fill=black!8, font=\scriptsize\bfseries},
  mj/group/.style={
    draw=black!40, rectangle, rounded corners=3pt, dashed,
    inner sep=2.5mm, fill=black!3
  },
  mj/label/.style={font=\small\itshape, anchor=south},
  mj/arr/.style={-{Latex[length=2mm]}, thick, black!70},
  mj/darr/.style={-{Latex[length=2mm]}, thick, black!55, dashed}
}

\usepackage{tcolorbox}
\usepackage{listings}
\usepackage{placeins} %

\usepackage{amsthm}
\theoremstyle{plain}

\theoremstyle{definition}

\theoremstyle{remark}

\newcommand{\parhead}[1]{\medskip \noindent \textbf{#1}\hskip .1in}

\newcommand{\SystemName}{\textsc{Mjolnir}\xspace}

\xspaceaddexceptions{\%}

\makeatletter
\newcommand{\TODO}[1]{{\let\xspace\relax\setlength{\fboxsep}{0pt}\TODO@words#1 \@nil}}
\def\TODO@words#1 #2\@nil{%
    \colorbox{yellow}{\strut#1}%
    \if\relax\detokenize{#2}\relax\expandafter\@gobble\else\expandafter\@firstofone\fi
    {\TODO@space\TODO@words#2\@nil}}
\def\TODO@space{\leaders\hbox{{\color{yellow}\vrule height\ht\strutbox depth\dp\strutbox}}%
    \hskip\fontdimen2\font plus\fontdimen3\font minus\fontdimen4\font}
\makeatother

\newcommand{\tgtCaliptraSwFiles}{64\xspace}
\newcommand{\tgtCaliptraSwCode}{12.5k\xspace}
\newcommand{\tgtCaliptraMcuSwFiles}{64\xspace}
\newcommand{\tgtCaliptraMcuSwCode}{12.8k\xspace}
\newcommand{\tgtCaliptraDpeFiles}{37\xspace}
\newcommand{\tgtCaliptraDpeCode}{14.3k\xspace}
\newcommand{\tgtOpentitanFwFiles}{23\xspace}
\newcommand{\tgtOpentitanFwCode}{1.3k\xspace}
\newcommand{\tgtOpentitanRtlFiles}{100\xspace}
\newcommand{\tgtOpentitanRtlCode}{23.5k\xspace}
\newcommand{\tgtOpenprotFiles}{56\xspace}
\newcommand{\tgtOpenprotCode}{7.8k\xspace}

\newcommand{\tgtCaliptraCoreFiles}{201\xspace}
\newcommand{\tgtCaliptraCoreCode}{37.0k\xspace}
\newcommand{\tgtTotalFiles}{344\xspace}
\newcommand{\tgtTotalCode}{72.2k\xspace}

\newcommand{\locPipeline}{2{,}837\xspace}
\newcommand{\locAgents}{1{,}705\xspace}
\newcommand{\locTools}{1{,}699\xspace}
\newcommand{\locPrompts}{245\xspace}
\newcommand{\locNix}{697\xspace}
\newcommand{\numAgents}{6\xspace}
\newcommand{\numTools}{12\xspace}

\newcommand{\cfgCaliptraSwProject}{12\xspace}

\newcommand{\cfgCaliptraSwShell}{28\xspace}

\newcommand{\cfgOpentitanProject}{55\xspace}
\newcommand{\cfgOpentitanJobs}{29\xspace}
\newcommand{\cfgOpentitanNumJobs}{7\xspace}
\newcommand{\cfgOpentitanShell}{177\xspace}

\newcommand{\cfgThreatModelMin}{59\xspace}
\newcommand{\cfgThreatModelMax}{84\xspace}

\newcommand{\priceFlashUncachedIn}{\$0.75\xspace}
\newcommand{\priceArgonUncachedIn}{\$2.00\xspace}
\newcommand{\evalRtOpenMean}{8\xspace}
\newcommand{\evalRtOpenMeanSd}{8 ($\pm$1.7)\xspace}
\newcommand{\evalRtPocVerifiedMeanSd}{7 ($\pm$2)\xspace}
\newcommand{\evalRtTotalInput}{336.3M\xspace}
\newcommand{\evalRtCachePct}{36.6\%\xspace}
\newcommand{\evalRtThinking}{2.47M\xspace}
\newcommand{\evalRtOutput}{0.44M\xspace}
\newcommand{\evalRtCost}{\$180.03\xspace}
\newcommand{\evalRtTime}{120.8\,min\xspace}
\newcommand{\evalRtTp}{24\xspace}
\newcommand{\evalRtFp}{0\xspace}
\newcommand{\evalRtPrecision}{100\%\xspace}
\newcommand{\evalRtLlmCalls}{6{,}424\xspace}
\newcommand{\evalRtToolCalls}{6{,}151\xspace}
\newcommand{\evalRtFileToolCalls}{3{,}122\xspace}
\newcommand{\evalRtSearchToolCalls}{2{,}690\xspace}
\newcommand{\evalRtAuditorToolCalls}{3{,}405\xspace}
\newcommand{\evalRtAuditorCost}{\$122.99\xspace}
\newcommand{\evalRtAuditorCostPct}{68.3\%\xspace}
\newcommand{\evalRtReviewerToolCalls}{946\xspace}
\newcommand{\evalRtRefusalsTotal}{1\xspace}
\newcommand{\evalRtRefusalsHard}{1\xspace}
\newcommand{\evalRtRefusalsSoft}{0\xspace}
\newcommand{\evalRtRefusalsApi}{0\xspace}
\newcommand{\evalRtSchemaDrops}{0\xspace}
\newcommand{\evalRtInitRevDropPct}{71.4\%\xspace}
\newcommand{\evalRtNoTmOpenMeanSd}{7 ($\pm$3)\xspace}
\newcommand{\evalRtNoTmPocVerifiedMeanSd}{7 ($\pm$3)\xspace}
\newcommand{\evalRtNoTmTotalInput}{294.7M\xspace}
\newcommand{\evalRtNoTmCachePct}{34.8\%\xspace}
\newcommand{\evalRtNoTmThinking}{2.30M\xspace}
\newcommand{\evalRtNoTmOutput}{0.42M\xspace}
\newcommand{\evalRtNoTmCost}{\$162.02\xspace}
\newcommand{\evalRtNoTmTime}{131.1\,min\xspace}
\newcommand{\evalRtNoTmTp}{16\xspace}
\newcommand{\evalRtNoTmFp}{5\xspace}
\newcommand{\evalRtNoTmPrecision}{76.2\%\xspace}
\newcommand{\evalRtNoTmToolCalls}{6{,}021\xspace}
\newcommand{\evalRtNoTmRefusalsTotal}{0\xspace}
\newcommand{\evalRtNoTmRefusalsHard}{0\xspace}
\newcommand{\evalRtNoTmRefusalsSoft}{0\xspace}
\newcommand{\evalRtNoTmRefusalsApi}{0\xspace}
\newcommand{\evalRtNoTmSchemaDrops}{0\xspace}
\newcommand{\evalRtNoExpertOpenMeanSd}{8.3 ($\pm$1.5)\xspace}
\newcommand{\evalRtNoExpertPocVerifiedMeanSd}{6.7 ($\pm$1.1)\xspace}
\newcommand{\evalRtNoExpertTotalInput}{300.3M\xspace}
\newcommand{\evalRtNoExpertCachePct}{10.7\%\xspace}
\newcommand{\evalRtNoExpertThinking}{2.61M\xspace}
\newcommand{\evalRtNoExpertOutput}{0.33M\xspace}
\newcommand{\evalRtNoExpertCost}{\$214.57\xspace}
\newcommand{\evalRtNoExpertTime}{155.6\,min\xspace}
\newcommand{\evalRtNoExpertTp}{20\xspace}
\newcommand{\evalRtNoExpertFp}{5\xspace}
\newcommand{\evalRtNoExpertPrecision}{80\%\xspace}
\newcommand{\evalRtNoExpertToolCalls}{5{,}854\xspace}
\newcommand{\evalRtNoExpertAuditorToolCalls}{4{,}056\xspace}
\newcommand{\evalRtNoExpertReviewerToolCalls}{1{,}335\xspace}
\newcommand{\evalRtNoExpertRefusalsTotal}{4\xspace}
\newcommand{\evalRtNoExpertRefusalsHard}{2\xspace}
\newcommand{\evalRtNoExpertRefusalsSoft}{1\xspace}
\newcommand{\evalRtNoExpertRefusalsApi}{1\xspace}
\newcommand{\evalRtNoExpertSchemaDrops}{0\xspace}
\newcommand{\evalRtPerFileNoTmOpenMeanSd}{26.3 ($\pm$2.1)\xspace}
\newcommand{\evalRtPerFileNoTmTotalInput}{0.24M\xspace}
\newcommand{\evalRtPerFileNoTmCachePct}{3.4\%\xspace}
\newcommand{\evalRtPerFileNoTmThinking}{1.09M\xspace}
\newcommand{\evalRtPerFileNoTmOutput}{0.01M\xspace}
\newcommand{\evalRtPerFileNoTmCost}{\$4.29\xspace}
\newcommand{\evalRtPerFileNoTmTime}{7.3\,min\xspace}
\newcommand{\evalRtPerFileNoTmTp}{9\xspace}
\newcommand{\evalRtPerFileNoTmFp}{3\xspace}
\newcommand{\evalRtPerFileNoTmPrecision}{75\%\xspace}
\newcommand{\evalRtPerFileNoTmToolCalls}{0\xspace}
\newcommand{\evalRtPerFileNoTmRefusalsTotal}{0\xspace}
\newcommand{\evalRtPerFileNoTmRefusalsHard}{0\xspace}
\newcommand{\evalRtPerFileNoTmRefusalsSoft}{0\xspace}
\newcommand{\evalRtPerFileNoTmRefusalsApi}{0\xspace}
\newcommand{\evalRtPerFileNoTmSchemaDrops}{0\xspace}
\newcommand{\evalRtPerFileOpenMeanSd}{33 ($\pm$3)\xspace}
\newcommand{\evalRtPerFileTotalInput}{0.31M\xspace}
\newcommand{\evalRtPerFileCachePct}{10.7\%\xspace}
\newcommand{\evalRtPerFileThinking}{1.16M\xspace}
\newcommand{\evalRtPerFileOutput}{0.01M\xspace}
\newcommand{\evalRtPerFileCost}{\$4.60\xspace}
\newcommand{\evalRtPerFileTime}{13.9\,min\xspace}
\newcommand{\evalRtPerFileTp}{9\xspace}
\newcommand{\evalRtPerFileFp}{3\xspace}
\newcommand{\evalRtPerFilePrecision}{75\%\xspace}
\newcommand{\evalRtPerFileToolCalls}{0\xspace}
\newcommand{\evalRtPerFileRefusalsTotal}{0\xspace}
\newcommand{\evalRtPerFileRefusalsHard}{0\xspace}
\newcommand{\evalRtPerFileRefusalsSoft}{0\xspace}
\newcommand{\evalRtPerFileRefusalsApi}{0\xspace}
\newcommand{\evalRtPerFileSchemaDrops}{0\xspace}
\newcommand{\evalRtWholeDirNoTmOpenMeanSd}{5 ($\pm$1.7)\xspace}
\newcommand{\evalRtWholeDirNoTmTotalInput}{26M\xspace}
\newcommand{\evalRtWholeDirNoTmCachePct}{91\%\xspace}
\newcommand{\evalRtWholeDirNoTmThinking}{0.06M\xspace}
\newcommand{\evalRtWholeDirNoTmOutput}{0.01M\xspace}
\newcommand{\evalRtWholeDirNoTmCost}{\$3.79\xspace}
\newcommand{\evalRtWholeDirNoTmTime}{23.7\,min\xspace}
\newcommand{\evalRtWholeDirNoTmTp}{10\xspace}
\newcommand{\evalRtWholeDirNoTmFp}{5\xspace}
\newcommand{\evalRtWholeDirNoTmPrecision}{66.7\%\xspace}
\newcommand{\evalRtWholeDirNoTmToolCalls}{180\xspace}
\newcommand{\evalRtWholeDirNoTmRefusalsTotal}{0\xspace}
\newcommand{\evalRtWholeDirNoTmRefusalsHard}{0\xspace}
\newcommand{\evalRtWholeDirNoTmRefusalsSoft}{0\xspace}
\newcommand{\evalRtWholeDirNoTmRefusalsApi}{0\xspace}
\newcommand{\evalRtWholeDirNoTmSchemaDrops}{0\xspace}
\newcommand{\evalRtWholeDirOpenMeanSd}{4.7 ($\pm$0.6)\xspace}
\newcommand{\evalRtWholeDirTotalInput}{27.8M\xspace}
\newcommand{\evalRtWholeDirCachePct}{94\%\xspace}
\newcommand{\evalRtWholeDirThinking}{0.07M\xspace}
\newcommand{\evalRtWholeDirOutput}{0.01M\xspace}
\newcommand{\evalRtWholeDirCost}{\$3.50\xspace}
\newcommand{\evalRtWholeDirTime}{30.1\,min\xspace}
\newcommand{\evalRtWholeDirTp}{11\xspace}
\newcommand{\evalRtWholeDirFp}{3\xspace}
\newcommand{\evalRtWholeDirPrecision}{78.6\%\xspace}
\newcommand{\evalRtWholeDirToolCalls}{201\xspace}
\newcommand{\evalRtWholeDirRefusalsTotal}{0\xspace}
\newcommand{\evalRtWholeDirRefusalsHard}{0\xspace}
\newcommand{\evalRtWholeDirRefusalsSoft}{0\xspace}
\newcommand{\evalRtWholeDirRefusalsApi}{0\xspace}
\newcommand{\evalRtWholeDirSchemaDrops}{0\xspace}
\newcommand{\evalRtArgonOpenMean}{22.7\xspace}
\newcommand{\evalRtArgonOpenMeanSd}{22.7 ($\pm$4.9)\xspace}
\newcommand{\evalRtArgonPocVerifiedMeanSd}{15 ($\pm$1.7)\xspace}
\newcommand{\evalRtArgonOpenUnverified}{7.7\xspace}
\newcommand{\evalRtArgonHarnessLimited}{6.7\xspace}
\newcommand{\evalRtArgonTotalInput}{62M\xspace}
\newcommand{\evalRtArgonCachePct}{84.9\%\xspace}
\newcommand{\evalRtArgonThinking}{0.34M\xspace}
\newcommand{\evalRtArgonOutput}{0.39M\xspace}
\newcommand{\evalRtArgonCost}{\$31.16\xspace}
\newcommand{\evalRtArgonTime}{112.8\,min\xspace}
\newcommand{\evalRtArgonSampled}{12\xspace}
\newcommand{\evalRtArgonTp}{10\xspace}
\newcommand{\evalRtArgonFp}{2\xspace}
\newcommand{\evalRtArgonPrecision}{83.3\%\xspace}
\newcommand{\evalRtArgonLlmCalls}{1{,}656\xspace}
\newcommand{\evalRtArgonToolCalls}{2{,}050\xspace}
\newcommand{\evalRtArgonAuditorCostPct}{36.4\%\xspace}
\newcommand{\evalRtArgonRefusalsTotal}{2\xspace}
\newcommand{\evalRtArgonRefusalsHard}{0\xspace}
\newcommand{\evalRtArgonRefusalsSoft}{1\xspace}
\newcommand{\evalRtArgonRefusalsApi}{1\xspace}
\newcommand{\evalRtArgonSchemaDrops}{0\xspace}
\newcommand{\evalRtArgonInitRevDropPct}{46.9\%\xspace}
\newcommand{\evalRtSonnetOpenMeanSd}{1 ($\pm$1)\xspace}
\newcommand{\evalRtSonnetPocVerifiedMeanSd}{1 ($\pm$1)\xspace}
\newcommand{\evalRtSonnetTotalInput}{8.32M\xspace}
\newcommand{\evalRtSonnetCachePct}{72.4\%\xspace}
\newcommand{\evalRtSonnetThinking}{0.05M\xspace}
\newcommand{\evalRtSonnetOutput}{0.13M\xspace}
\newcommand{\evalRtSonnetCost}{\$7.62\xspace}
\newcommand{\evalRtSonnetTime}{5.9\,min\xspace}

\newcommand{\evalRtSonnetTp}{3\xspace}
\newcommand{\evalRtSonnetFp}{0\xspace}
\newcommand{\evalRtSonnetPrecision}{100\%\xspace}
\newcommand{\evalRtSonnetToolCalls}{293\xspace}
\newcommand{\evalRtSonnetRefusalsTotal}{0\xspace}
\newcommand{\evalRtSonnetRefusalsHard}{0\xspace}
\newcommand{\evalRtSonnetRefusalsSoft}{0\xspace}
\newcommand{\evalRtSonnetRefusalsApi}{0\xspace}
\newcommand{\evalRtSonnetSchemaDrops}{8\xspace}
\newcommand{\evalRtSonnetInitRevDropPct}{97\%\xspace}
\newcommand{\evalRtFlashArgonInputRatio}{5.4$\times$\xspace}
\newcommand{\evalRtStabilityOpenMeanSd}{9 ($\pm$2.5)\xspace}
\newcommand{\evalRtStabilityPocVerifiedMeanSd}{7.4 ($\pm$2.1)\xspace}
\newcommand{\evalRtStabilityDiscOpen}{30\xspace}
\newcommand{\evalCaliptraSwPoolInSample}{138\xspace}
\newcommand{\evalCaliptraSwPoolDoubleReviewed}{56\xspace}
\newcommand{\evalCaliptraMcuSwPocVerified}{8\xspace}
\newcommand{\evalCaliptraMcuSwDisprovedByHarness}{1\xspace}
\newcommand{\evalCaliptraMcuSwCost}{\$113.89\xspace}
\newcommand{\evalCaliptraMcuSwRefusalsTotal}{1\xspace}
\newcommand{\evalCaliptraMcuSwRefusalsHard}{0\xspace}
\newcommand{\evalCaliptraMcuSwRefusalsSoft}{1\xspace}
\newcommand{\evalCaliptraMcuSwRefusalsApi}{0\xspace}
\newcommand{\evalCaliptraMcuSwSchemaDrops}{0\xspace}
\newcommand{\evalCaliptraMcuSwFinalOpen}{10\xspace}
\newcommand{\evalCaliptraDpePocVerified}{0\xspace}
\newcommand{\evalCaliptraDpeCost}{\$39.36\xspace}
\newcommand{\evalCaliptraDpeRefusalsTotal}{0\xspace}
\newcommand{\evalCaliptraDpeRefusalsHard}{0\xspace}
\newcommand{\evalCaliptraDpeRefusalsSoft}{0\xspace}
\newcommand{\evalCaliptraDpeRefusalsApi}{0\xspace}
\newcommand{\evalCaliptraDpeSchemaDrops}{0\xspace}
\newcommand{\evalCaliptraDpeDiscOpen}{7\xspace}
\newcommand{\evalCaliptraDpeFinalOpen}{0\xspace}
\newcommand{\evalOpentitanFwPocVerified}{2\xspace}
\newcommand{\evalOpentitanFwCost}{\$47.34\xspace}
\newcommand{\evalOpentitanFwRefusalsTotal}{0\xspace}
\newcommand{\evalOpentitanFwRefusalsHard}{0\xspace}
\newcommand{\evalOpentitanFwRefusalsSoft}{0\xspace}
\newcommand{\evalOpentitanFwRefusalsApi}{0\xspace}
\newcommand{\evalOpentitanFwSchemaDrops}{0\xspace}
\newcommand{\evalOpentitanFwFinalOpen}{3\xspace}
\newcommand{\evalOpentitanRtlPocVerified}{5\xspace}
\newcommand{\evalOpentitanRtlCost}{\$146.16\xspace}
\newcommand{\evalOpentitanRtlRefusalsTotal}{0\xspace}
\newcommand{\evalOpentitanRtlRefusalsHard}{0\xspace}
\newcommand{\evalOpentitanRtlRefusalsSoft}{0\xspace}
\newcommand{\evalOpentitanRtlRefusalsApi}{0\xspace}
\newcommand{\evalOpentitanRtlSchemaDrops}{0\xspace}
\newcommand{\evalOpentitanRtlFinalOpen}{5\xspace}
\newcommand{\evalOpenprotPocVerified}{8\xspace}
\newcommand{\evalOpenprotHarnessLimited}{3\xspace}
\newcommand{\evalOpenprotCost}{\$60.46\xspace}
\newcommand{\evalOpenprotRefusalsTotal}{0\xspace}
\newcommand{\evalOpenprotRefusalsHard}{0\xspace}
\newcommand{\evalOpenprotRefusalsSoft}{0\xspace}
\newcommand{\evalOpenprotRefusalsApi}{0\xspace}
\newcommand{\evalOpenprotSchemaDrops}{0\xspace}
\newcommand{\evalOpenprotFinalOpen}{11\xspace}
\newcommand{\evalRtStabilityExtraRefusalsTotal}{3\xspace}
\newcommand{\evalRtStabilityExtraRefusalsHard}{2\xspace}
\newcommand{\evalRtStabilityExtraRefusalsSoft}{0\xspace}
\newcommand{\evalRtStabilityExtraRefusalsApi}{1\xspace}
\newcommand{\evalRtStabilityExtraSchemaDrops}{0\xspace}
\newcommand{\evalTotalUniqueRuns}{34\xspace}
\newcommand{\evalTotalSafetyRefusals}{11\xspace}
\newcommand{\evalTotalHardSafetyRefusals}{5\xspace}
\newcommand{\evalTotalSoftRefusals}{3\xspace}
\newcommand{\evalTotalApiSafetyBlocks}{3\xspace}
\newcommand{\evalTotalSchemaValidationDrops}{8\xspace}
\newcommand{\evalKbFlashA}{4\xspace}
\newcommand{\evalKbFlashB}{6\xspace}
\newcommand{\evalKbFlashC}{0\xspace}
\newcommand{\evalKbFlashAll}{10\xspace}
\newcommand{\evalKbFlashDisc}{18\xspace}
\newcommand{\evalKbFlashClosed}{8\xspace}
\newcommand{\evalKbFlashPoc}{8\xspace}
\newcommand{\evalKbFlashAllPct}{29.4\%\xspace}
\newcommand{\evalKbFlashBaseA}{3\xspace}
\newcommand{\evalKbFlashBaseB}{4\xspace}
\newcommand{\evalKbFlashBaseC}{0\xspace}
\newcommand{\evalKbFlashBaseAll}{7\xspace}
\newcommand{\evalKbFlashBaseAllPct}{20.6\%\xspace}

\newcommand{\evalKbArgonA}{5\xspace}
\newcommand{\evalKbArgonB}{12\xspace}
\newcommand{\evalKbArgonC}{0\xspace}
\newcommand{\evalKbArgonAll}{17\xspace}
\newcommand{\evalKbArgonDisc}{19\xspace}
\newcommand{\evalKbArgonClosed}{2\xspace}
\newcommand{\evalKbArgonPoc}{17\xspace}
\newcommand{\evalKbArgonAllPct}{50\%\xspace}
\newcommand{\evalKbArgonBaseA}{4\xspace}
\newcommand{\evalKbArgonBaseB}{8\xspace}
\newcommand{\evalKbArgonBaseC}{0\xspace}
\newcommand{\evalKbArgonBaseAll}{12\xspace}
\newcommand{\evalKbArgonBaseAllPct}{35.3\%\xspace}

\newcommand{\overlapTolerance}{10\xspace}
\newcommand{\overlapStabIssues}{25\xspace}
\newcommand{\overlapStabAllRuns}{3\xspace}
\newcommand{\overlapStabMajority}{4\xspace}
\newcommand{\overlapStabTwo}{5\xspace}
\newcommand{\overlapStabMulti}{9\xspace}
\newcommand{\overlapStabSingle}{16\xspace}

\newcommand{\overlapStabRecurringPct}{67\%\xspace}
\newcommand{\overlapModelIssues}{44\xspace}
\newcommand{\overlapModelAll}{2\xspace}
\newcommand{\overlapModelTwo}{6\xspace}
\newcommand{\overlapModelOne}{36\xspace}
\newcommand{\overlapModelFlashIssues}{16\xspace}
\newcommand{\overlapModelArgonIssues}{35\xspace}
\newcommand{\overlapModelSonnetIssues}{3\xspace}

\newcommand{\evalConfirmedTotal}{64\xspace}
\newcommand{\evalConfirmedRt}{37\xspace}
\newcommand{\evalConfirmedOther}{27\xspace}

\newcommand{\callsExpert}{150\xspace}
\newcommand{\callsAuditor}{150\xspace}
\newcommand{\callsIngestor}{150\xspace}
\newcommand{\callsDeduplicator}{50\xspace}
\newcommand{\callsReviewer}{200\xspace}
\newcommand{\callsExploiter}{150\xspace}
\newcommand{\harnessTimeout}{900\,s\xspace}
\newcommand{\toolOutputMax}{40{,}000\xspace}

\ifdefined\isFinalized
\newcommand{\NewCommentType}[3]{}
\else
\newcommand{\NewCommentType}[3]{\expandafter\newcommand\csname #1\endcsname[1]{{\color{#2}{#3: ##1}} }}
\fi

\begin{document}
\date{}

\title{\Large \bf \SystemName: Finding and Reproducing Vulnerabilities in
Root-of-Trust Firmware and RTL with Project-Grounded Agents}

\author{
{\rm Collin MacDonald}\thanks{Work done while the author was at Google.}\\
William \& Mary\\
\texttt{cmacdonald01@wm.edu}
\and
{\rm Carl Lundin}\footnotemark[1]\\
Caliptra
\and
{\rm Stephen Herwig}\\
William \& Mary\\
\texttt{smherwig@wm.edu}
\and
{\rm Miguel Osorio}\footnotemark[1]\\
Ricursive Intelligence\\
\texttt{miguel@ricursive.com}
\and
{\rm Timothy Trippel}\footnotemark[1]\\
Ricursive Intelligence\\
\texttt{tim@ricursive.com}
} %

\begin{NoHyper}
\maketitle
\end{NoHyper}

\begin{abstract}
A root of trust (RoT) anchors secure boot and attestation for the system above
it, so a bug in RoT firmware or hardware undermines every layer that trusts it.
LLM agents now find real vulnerabilities in general-purpose software, but RoT
code presents two challenges.
First, many security-relevant bugs depend on project-specific trust boundaries
and fault-injection countermeasures rather than generic vulnerability patterns,
and may cause no crash.
Second, without RoT hardware, agents must exercise RoT inputs through the
project's own hardware model, mocks, or RTL testbenches.

We present \SystemName, an agentic pipeline that supplies each project's threat
model and design conventions to discovery and skeptical review agents, and
reproduces surviving findings as proof-of-concept tests in the project's own
toolchain.
We evaluate \SystemName on six firmware and RTL targets in Caliptra, OpenTitan,
and OpenPRoT\@.
On Caliptra's runtime firmware, it reaches 83.3\%--100\% precision across three
models under blind expert labeling, including 100\% on Gemini~3.8 Flash versus
66.7\%--78.6\% for four single-agent baselines on the same model.
On 34 known security-relevant bugs in Caliptra's core firmware fixed after the
models' training cutoffs, it finds 10 with Gemini~3.8 Flash versus 7 for the
best baseline, and 17 with Gemini~4 Argon versus 12.
Across our evaluation scans, maintainers have confirmed 64 of \SystemName's
findings as previously unknown security-relevant bugs: 37 across three models on
Caliptra's runtime firmware and 27 across single scans of the other five targets.

\end{abstract}

\section{Introduction}
\label{sec:intro}

A bug in a root of trust (RoT) can undermine every system that relies on it,
and some RoT bugs can never be patched. An RoT, either a discrete chip or a
block within a system-on-chip (SoC), provides secure boot and attestation to
the system around it.
Open-source RoTs now ship in production: Caliptra provides an RoT for
datacenter SoCs~\cite{caliptra_spec_2025}, OpenTitan ships in
Chromebooks~\cite{opentitan_history_2026}, and OpenPRoT provides firmware for
platform RoTs~\cite{openprot_docs_2026}.
A flaw in RoT firmware or hardware can let an attacker bypass secure boot,
extract secret keys, or forge attestation evidence.
Flaws in the boot ROM or in the register-transfer level (RTL) hardware design
are fixed at fabrication, so they persist in every deployed
device~\cite{cert_vu941987_2019, dessouky_hardfails_2019, kumar_nintendo_2018,
opentitan_rom_2026}.
RoT bugs must therefore be found before release.

LLM agents now find real vulnerabilities in general-purpose
software~\cite{big_sleep_team_naptime_2024}, but RoT firmware and RTL challenge
the assumptions underlying existing approaches.
Systems in DARPA's AI Cyber Challenge (AIxCC) confirm bugs through crashes and
sanitizer reports~\cite{aixcc_2026, zhang_sok_2026}, while OpenAnt and
Co-RedTeam reproduce vulnerabilities using proof-of-concept (PoC) exploits in
their own containers~\cite{korda_openant_2026, he_co-redteam_2026}.
Neither approach suffices for RoT code for two reasons.

First, whether RoT code is buggy depends on the project's trust boundaries,
which generic vulnerability patterns do not encode.
For example, Caliptra's runtime firmware accepted certain mailbox commands
without checking the caller's privilege level or the bus address it supplied,
which enabled denial of service or arbitrary direct memory access
(DMA)~\cite{caliptra_cve_7328_2026}.
The bug involves no memory corruption.
It violates Caliptra's trust boundary because it lets unprivileged callers
issue commands reserved for privileged callers. We call a bug
\emph{security-relevant} if it lets adversary-controlled input violate one of
the project's security objectives or if it omits a countermeasure that the
project's threat model requires.
Of the 34 known security-relevant bugs we study in Caliptra's core firmware, 15
fall under CWEs for missing authorization, improper authentication, improper
signature verification, or time-of-check-to-time-of-use races.
To find such bugs, an agent must know which callers and inputs each part of the
RoT trusts.

Beyond trust boundaries, RoT code also contains countermeasures that resemble
bugs, so an agent without the project's design context errs in two directions.
It flags intentional countermeasures as redundant code, a false positive, and
it overlooks security checks that lack a required countermeasure, a false
negative.
OpenTitan's first mutable boot stage re-checks values immediately after a
switch matches them, and it encodes booleans as \texttt{0x739} and
\texttt{0x1d4} rather than \texttt{1} and
\texttt{0}~\cite{opentitan_rom_ext_2026, opentitan_hardened_2026}.
These redundancies are designed to prevent a single glitched branch or flipped
bit from bypassing a security check.
To judge RoT code correctly, an agent must also know which countermeasures the
project requires.

Second, an agent without RoT hardware can exercise adversary-controlled inputs
only through the project's own test infrastructure.
These inputs include mailbox commands, firmware images, and platform management
messages, which generic HTTP, command-line, or file-based harnesses cannot
supply~\cite{korda_openant_2026}.
A PoC must instead drive Caliptra's hardware model, OpenTitan's Bazel tests, or
a Verilator RTL testbench.
It must also check a security property rather than wait for a crash, because a
logic flaw need not cause a crash or trigger the sanitizer checks that AIxCC
systems rely on~\cite{zhang_sok_2026}.
A PoC that replaces the target logic with ad hoc stubs demonstrates the
behavior of those stubs, not a flaw in the project's implementation.
Maintainers already spend effort triaging unverified
reports~\cite{baltes_endless_2026}, so a useful PoC must run as a test in the
project's own harness, where maintainers can inspect and rerun it.

Our insight is to ground every stage of vulnerability discovery, review, and
reproduction in the RoT project's own artifacts.
We keep these stages from prior agentic systems such as
OpenAnt~\cite{korda_openant_2026} and Co-RedTeam~\cite{he_co-redteam_2026} and
change the evidence each stage uses.
Discovery and review judge findings against the project's threat model and
design conventions, and reproduction runs PoCs on the project's hardware model,
mocks, or testbenches instead of an environment the agent builds.

We present \SystemName,\footnote{\url{https://github.com/chipsalliance/mjolnir}}{}
an agentic pipeline that applies this insight: it grounds each stage in project
context and reproduces surviving findings in the project's own harness.
Before scanning, a \texttt{Project Expert} agent summarizes the project's
execution stages, trust boundaries, and intentional countermeasures, and it
later answers other agents' questions.
Every agent also receives a per-project threat model of at most 90 lines.
\texttt{Source File Auditor} agents propose candidate findings in parallel, a
\texttt{Deduplicator} agent merges findings that share a root cause, and a
skeptical \texttt{Adversarial Reviewer} agent rejects findings that the source
code and threat model do not support.
A \texttt{PoC Exploit Generator} agent then reproduces surviving findings as
tests in the project's toolchain.

\SystemName checks whether each PoC triggers the reported flaw through
unmodified production code, consistent with an adversary who controls the RoT's
inputs but not its code.
An agent might otherwise manufacture a PoC by weakening a security check or
mocking the code under test, without demonstrating a flaw in the original
implementation.
\SystemName therefore checks each PoC twice.
An automated diff check marks a PoC unverified unless it ran a harness command
and changed only test files, and a final adversarial review rejects PoCs that
mock the code under test or make vacuous assertions, which pass whether or not
the flaw exists.
A failed or rejected PoC does not by itself refute a finding.
\SystemName keeps the finding open as unverified unless the attempt reveals
evidence that the reported flaw is not exploitable.
Onboarding a project takes four configuration files and no changes to
\SystemName, and the same agents audit SystemVerilog RTL with language-specific
guidance.

We evaluate \SystemName on six targets across Caliptra, OpenTitan, and
OpenPRoT, covering C and Rust firmware and SystemVerilog RTL\@.
Because the complete set of bugs in a real codebase is unknown, we cannot
measure recall on previously unknown bugs.
We therefore combine three sources of evidence.
First, Caliptra maintainers blindly label open findings on Caliptra's runtime
firmware.
There, \SystemName reaches \evalRtArgonPrecision--\evalRtPrecision precision
across three models, and \evalRtPrecision on Gemini~3.8 Flash (\evalRtTp of
\evalRtTp findings) compared to
\evalRtWholeDirNoTmPrecision--\evalRtWholeDirPrecision for four single-agent
baselines on the same model.
Second, we use 34 known bugs in Caliptra's core firmware, all fixed after the
models' published training cutoffs.
\SystemName finds 10 of them with Gemini~3.8 Flash and 17 with Gemini~4 Argon,
compared to 7 and 12 for the best baseline on each model.
Third, across our evaluation scans, maintainers have confirmed
\evalConfirmedTotal of \SystemName's findings as previously unknown
security-relevant bugs (\evalConfirmedRt across the three models' runs on
Caliptra's runtime firmware and \evalConfirmedOther across single scans of the
other five targets). The evaluation also shows limits.
With Gemini~3.8 Flash, \SystemName's review stages wrongly dismiss 8 of the 18
known bugs that its auditors find, labeling them false positives or not
exploitable: 7 in initial review, which receive no PoC or further analysis, and
1 in final review.
Across five Flash runs, 16 of 25 location-matched issues appear in only one
run, so repeated scans find more bugs than a single scan.

We make the following contributions:
\begin{itemize}
\item \textbf{Project-grounded discovery and reproduction.}
    We design \SystemName, which supplies each project's threat model and design
    conventions to discovery and skeptical review, and reproduces surviving
    findings as tests in the project's own toolchain. 
    A diff check and final review reject PoCs that edit production code or mock
    the target (\S\ref{sec:design}--\S\ref{sec:implementation}).

\item \textbf{Evaluation on RoT firmware and RTL.}
    We evaluate \SystemName on six targets against four single-agent baselines,
    two ablations, and three models.
    It includes 34 known security-relevant bugs in Caliptra's core firmware
    fixed after the models' training cutoffs (\S\ref{sec:evaluation},
    Appendix~\ref{sec:appendix-known-bugs}).

\item \textbf{Real-world findings and deployment.}
    Maintainers have confirmed \evalConfirmedTotal of \SystemName's findings as
    previously unknown security-relevant bugs.
    \SystemName also runs on pull requests in three RoT repositories, where it
    has scanned over 800 pull requests in four months
    (\S\ref{sec:impl-deployment}).
\end{itemize}

\section{Background and Threat Model}
\label{sec:background}

\subsection{Root-of-Trust Firmware and Hardware}
\label{sec:bg-rot}

An RoT establishes trust in a system through secure boot and attestation. In
secure boot, immutable ROM code verifies a signature over the next firmware
stage with a public key fixed in ROM, fuses, or one-time-programmable (OTP)
memory, and each stage then verifies the next~\cite{caliptra_spec_2025,
21-sec-dice}.
In measured boot, each stage also hashes the next stage before running it.
The Device Identifier Composition Engine (DICE) uses a Unique Device Secret
(UDS) and firmware measurements to derive a stage-specific secret called a
Compound Device Identifier (CDI)~\cite{21-sec-dice, tcg_dice_layering_2020}.
Modified firmware therefore receives different derived keys.
For attestation, the RoT signs its measurements using CDI-derived keys, which a
remote verifier checks~\cite{tcg_dice_layering_2020}.
An RoT must therefore keep its secrets confidential and must never run or
attest to unauthorized firmware.

\parhead{Caliptra.}
Caliptra is an RoT for datacenter SoCs whose Rust firmware boots in three
stages: an immutable ROM, the First Mutable Code (FMC), and the
Runtime~\cite{caliptra-sw_2026, caliptra_spec_2025}.
The ROM validates the signed firmware manifest and checks the FMC and
Runtime images against it, and the FMC derives the Runtime's DICE identity.
The Runtime implements over 100 mailbox commands that the SoC issues over
its AXI bus.
Caliptra's hardware controls access to the mailbox, but the
firmware defines what each command does and checks whether the caller may
issue it~\cite{caliptra_spec_2025}.
Some commands invoke the DICE Protection Environment (DPE), which derives
identities for other SoC components~\cite{caliptra-dpe_2026}.
Caliptra 2.0 adds a Manufacturer Control Unit (MCU), whose ROM brings up the
firmware recovery interface and whose firmware receives SoC firmware updates
over the Management Component Transport Protocol (MCTP) and Platform Level Data
Model (PLDM)~\cite{caliptra-mcu-sw_2026, caliptra_spec_2025}.

\parhead{OpenTitan.}
OpenTitan's Earl Grey is a discrete secure microcontroller with a RISC-V
core~\cite{opentitan_earlgrey_2026}.
Its ROM, fixed at fabrication, authenticates the first mutable boot stage,
ROM\_EXT, against the Silicon Creator's public keys in OTP
memory~\cite{opentitan_rom_2026, opentitan_root_keys_2026}.
ROM\_EXT authenticates the owner's firmware and restricts its access to
security-critical device assets before handing it
control~\cite{opentitan_rom_ext_doc_2026}.
Through boot-services requests, owner code can ask ROM\_EXT to change the next
boot slot, the minimum firmware security version, or the device's
owner~\cite{opentitan_secure_boot_2026, opentitan_boot_svc_2026}.
Our RTL target is Earl Grey's top level, which connects the chip's IP blocks
and contains chip-specific blocks such as the flash controller and alert
handler~\cite{opentitan_earlgrey_2026}.

\parhead{OpenPRoT.}
OpenPRoT is an open-source firmware stack for platform RoTs, which verify the
firmware of a server's other components before releasing them from
reset~\cite{openprot_spec_2026, openprot_orchestrator_2026}.
Its services run on Pigweed's \texttt{pw\_kernel} microkernel and build with
Bazel~\cite{openprot_pigweed_2026, pigweed_2026}.
They exchange MCTP messages, handle attestation over the Security Protocol and
Data Model (SPDM), and accept PLDM firmware updates~\cite{pldm_fwup_2023,
spdm_2024, mctp_2026}.

\subsection{LLM Agents}
\label{sec:bg-agents}

An LLM agent alternates between model calls and tool calls, using results such
as file contents or test output to guide subsequent
steps~\cite{yang_swe-agent_2024, 23-iclr-react}.
\SystemName uses Google's Agent Development Kit (ADK) to compose agents with
separate instructions, tools, and output schemas into
workflows~\cite{adk_2026}.
Because model outputs are stochastic, repeated scans of the same code can
produce different findings.

\subsection{Threat Model and Analysis Scope}
\label{sec:threat-model}

\parhead{Adversary.}
The adversary seeks to violate the guarantees that an RoT provides to the
system above it, and controls every input that crosses the RoT's external
interfaces.
In Caliptra, these inputs are mailbox commands, firmware images, fuse values,
and the firmware recovery interface, plus MCTP and PLDM messages to the
MCU~\cite{caliptra_spec_2025}.
In OpenTitan, they are boot images, boot-services
requests, and ownership changes.
In OpenPRoT, they are MCTP, SPDM, and PLDM messages from other platform
components. The projects' own threat models identify the assets to protect:
secure boot and anti-rollback, the secrecy of device secrets and derived keys
such as the UDS and CDIs, and the soundness of
attestation~\cite{opentitan_threat_model_2026, caliptra_spec_2025,
openprot_threat_model_2026}.
For each target, we give \SystemName's agents a threat model that extends the
project's own with one more security objective: the availability of RoT
services.

\parhead{Security-relevant bugs.}
We call a bug \emph{security-relevant} if it lets adversary-controlled input
violate one of the project's security objectives, or if it omits a
countermeasure that the project's threat model requires.
The security objectives are properties the RoT must enforce: the integrity of
secure boot and anti-rollback, the secrecy of device secrets and keys,
privilege separation across the RoT's trust boundaries, and the
availability of RoT services across resets.
A \emph{countermeasure} is code that the threat model requires to limit the
damage of an attack, such as a hardened comparison that resists fault injection
or the zeroization of a secret after use.
A bug \emph{omits} a countermeasure if the required code is missing, or if it
protects some paths to a security decision but not others.
We label bugs of this second kind \emph{defense in depth}, because exploiting
them requires an additional capability, such as a physical attack.
Neither kind requires a demonstrated attack on deployed hardware: a bug counts
if the code fails to enforce an objective or a required countermeasure, and
expert labels and the known-bug set count both kinds.

\parhead{Physical attacks and assumptions.}
All three projects place physical attacks, such as fault injection and side
channels, in scope and require countermeasures against them~\cite{opentitan_life_cycle_2026,
caliptra_spec_2025, openprot_threat_model_2026}.
\SystemName does not mount these attacks, but it reports code that omits a
required countermeasure.
We assume that cryptographic algorithms are secure and that hardware outside
each target behaves according to its specification.

\parhead{Evidence standard.}
We consider a PoC valid evidence of a finding only if it triggers the reported
flaw through adversary-controlled inputs to unmodified production code.
A PoC may construct inputs, drive the code through the entry points an attacker
reaches, and reuse the project's existing mocks for hardware outside the
target.
It may not change production code or replace the code under test, because the
adversary can do neither.
A PoC that breaks this rule shows only that modified code is buggy, leaving the
finding unverified rather than false.

\parhead{Operational assumptions.}
A project's maintainers or security reviewers run \SystemName on demand or on
each pull request.
\SystemName sends the code and threat model to a model provider that we assume
is trusted, while projects that cannot share code can run open-weight models
locally.
We assume that the code under analysis does not attempt to manipulate
\SystemName, for example through prompt injections in comments.
This assumption may not hold for pull requests from untrusted contributors, so
\SystemName supplements rather than replaces human review, which must also
consider attacks directed at the analyzer.
\SystemName runs PoCs only on development machines, as host-side tests or
Verilator testbenches, and never on RoT hardware.
In \texttt{Re-Analysis} mode, \SystemName also verifies external vulnerability
reports.
Because such reports may misstate the project or its threat model, their
findings pass through the same skeptical review and PoC stages as discovered
findings.

\section{System Design}
\label{sec:design}

\SystemName structures its analysis as a multi-stage funnel that
grounds discovery in project context, filters candidates through deduplication
and skeptical adversarial review, and synthesizes executable PoCs in the
project's own test harness for surviving findings.
Throughout the pipeline, agents combine a per-project threat model with scoped
code search tools and on-demand answers from a \texttt{Project Expert} agent.
Designed primarily for RoT firmware, the same pipeline extends without code
changes to SystemVerilog RTL.

\subsection{System Design Goals}
\label{sec:design-goals}

\SystemName's design goals follow from the requirements in
\S\ref{sec:intro}, and from the need to support RoT projects with different
languages and toolchains:

\begin{itemize}
    \item \textbf{D1: Project-grounded discovery.}
        Agents judge code against the project's architecture, trust
        boundaries, and threat model, not against the contents of a single
        file alone.
    \item \textbf{D2: Staged review and native reproduction.}
        Skeptical source-grounded review filters false positives before
        execution, and surviving findings are reproduced where possible as
        test-only PoCs in the project's own harness.
    \item \textbf{D3: Portability across RoT projects.}
        Onboarding a project requires only a Nix project definition, a threat
        model, scan jobs, and a development shell with the project's
        toolchain~\cite{nixos_flakes_2026}, without changing \SystemName.
\end{itemize}

\subsection{Pipeline Overview}
\label{sec:design-pipeline}

\begin{figure*}[t]
    \centering
    \begin{tikzpicture}[
    ag/.style={mj/agent, text width=19mm, minimum height=9mm},
    io/.style={mj/box, text width=15mm, minimum height=7mm},
  ]
  \node[io] (src) at (0.9, 0.6) {Source files};
  \node[io] (rep) at (0.9,-0.6) {External report};

  \node[ag] (aud) at (3.35, 0.6) {Source File Auditor};
  \node[ag] (ing) at (3.35,-0.6) {Report Ingestor};
  \node[ag] (dd)  at (5.8, 0)    {Deduplicator};
  \node[ag] (r1)  at (8.2, 0)    {Adversarial Reviewer (initial)};
  \node[ag] (poc) at (10.6,0)    {PoC Exploit Generator};
  \node[ag] (r2)  at (13.0,0)    {Adversarial Reviewer (final)};
  \node[mj/box, text width=18mm, minimum height=9mm] (out) at (15.4,0) {Verified findings};

  \node[ag] (pe) at (8.2, 1.9) {Project Expert};

  \node[mj/box, text width=34mm, minimum height=8mm, fill=black!15, font=\scriptsize\bfseries]
    (sb) at (10.6,-1.5) {\mbox{Project environment}\\{\normalfont Nix toolchain + test harness}};

  \draw[mj/arr] (src) -- (aud);
  \draw[mj/arr] (rep) -- (ing);
  \draw[mj/arr] (aud.east) -- (dd.west);
  \draw[mj/arr] (ing.east) -- (dd.west);
  \draw[mj/arr] (dd) -- (r1);
  \draw[mj/arr] (r1) -- (poc);
  \draw[mj/arr] (poc) -- (r2);
  \draw[mj/arr] (r2) -- (out);
  \draw[mj/arr, <->] (poc) -- (sb);

  \foreach \n in {aud, dd, r1, poc, r2}
    \draw[mj/darr] (pe) -- (\n.north);
\end{tikzpicture}
    \caption{\SystemName's pipeline.}
    \label{fig:design}
\end{figure*}

\SystemName orchestrates specialized agents alongside domain-specific tools, as
shown in Figure~\ref{fig:design}.
As input, \SystemName takes the project's source code, its threat model, and a
\emph{scan job}: a configuration file, written when the project is onboarded,
that names the model and selects the firmware or RTL files to analyze, such as
all SystemVerilog sources except generated register files.
We call the selected files \emph{in scope}.

The \texttt{Project Expert} agent first surveys the project and writes a summary
of its architecture, which every later agent receives.
The \texttt{Source File Auditor} agent then analyzes each in-scope file and
reports candidate findings, and the \texttt{Deduplicator} agent merges findings
that share a root cause.
The \texttt{Adversarial Reviewer} agent audits each remaining finding as part of
an initial review.
For each finding that survives, the \texttt{PoC Exploit Generator} agent writes
a PoC and runs it in the project's own test harness.
The \texttt{Adversarial Reviewer} agent then inspects the PoC and its output as
part of the final review process.
Findings are closed and removed from the pipeline once they are marked as a
duplicate, false positive, or non-exploitable result.
As a result, later stages such as PoC creation process fewer findings than
earlier ones.
\texttt{Re-Analysis} mode changes only the start of this pipeline.

Every stage reads and updates one record per finding.
The record holds the finding's file and location, severity, CWE, Common
Vulnerability Scoring System (CVSS) vector, verdict, status, and PoC.
\SystemName checks the CWE identifier against MITRE's catalog and recomputes
the CVSS score from the vector rather than trusting a score the model reports.
Each stage merges its output into the record and appends a snapshot to the
record's history, so every verdict can be traced to the stage that made it.
A False Positive or Not Exploitable verdict closes the finding, and the
\texttt{Deduplicator} agent marks each merged finding as a duplicate of a
canonical one.

\subsection{Agents}
\label{sec:design-agents}

Each agent has its own instructions, a fixed set of tools, and a bounded
number of model calls.
Every agent except the \texttt{Report Ingestor} agent, which reads external
vulnerability reports in \texttt{Re-Analysis} mode (described further in this
section), receives the project's threat model in its instructions.
Each threat model is a Markdown document of 90 lines or less that lists the
project's trust boundaries, assets, and key threat categories, such as fault
injection and privilege escalation.
These agents also receive the \texttt{Project Expert} agent's summary.
Every agent other than the \texttt{Project Expert} agent returns its result in
a fixed schema, which \SystemName merges into the finding record.
We describe the agents in pipeline order, except for the PoC Exploit Generator
agent, which we discuss alongside PoC reproduction and final review in
\S\ref{sec:design-verification}.

\parhead{\texttt{Project Expert} agent.}
The \texttt{Project Expert} agent surveys a project before auditing begins, and
other agents query it when they have project-specific questions (D1).
It runs first and surveys the project with read-only code search tools,
including its directory layout, READMEs, architecture documents, and build
files.
It then writes a summary of the project's execution stages, trust boundaries
and security invariants, register maps and memory-mapped I/O conventions, and
build and test targets.
The summary also lists intentional design patterns that could resemble bugs to
an unaware agent, such as OpenTitan's hardened checks.
\SystemName caches the summary per commit and threat model.

Other agents ask the \texttt{Project Expert} agent questions through a tool.
Each question starts a fresh instance of the agent, seeded with the threat
model, the summary, and the run's earlier questions and answers.
A repeated question is answered from that history without a new instance.
Because of this design, the \texttt{Project Expert} agent's context stays
bounded, even when many agents query it in parallel.

\parhead{\texttt{Source File Auditor} agent.}
The \texttt{Source File Auditor} agent is the entry point of \texttt{Discovery}
mode and finds candidate bugs in one source file at a time (D1).
\SystemName runs one Auditor per in-scope file in parallel, and gives each the
file's path and full contents.
No cross-file context is precomputed.
Instead, the Auditor retrieves callers, callees, and type definitions with code
search tools, and asks the \texttt{Project Expert} agent about project
conventions.
Its instructions bound this exploration to what it needs to confirm or refute a
suspected bug.

The Auditor's instructions also encode RoT-specific audit guidance.
They direct it to check integer arithmetic, untrusted inputs from peripherals
and host commands, concurrency, timing and power side channels in cryptographic
code, the zeroization of secrets, and security decisions that must resist fault
injection.
Language-specific guidance, selected by file type, covers memory safety,
\texttt{unsafe} code, error handling, and panics for C and Rust.
For SystemVerilog, it covers bit-width and signedness hazards, register-lock
bypass, clock- and reset-domain crossings, and state-machine integrity: a
hardware state machine must never enter an undefined state or skip a required
one.
The instructions also state false-positive constraints for embedded targets: an
in-order core without speculation or a data cache, a watchdog that resets the
chip on a crash, and an attacker who reaches the firmware only through its
communication interfaces.
As a result, the Auditor reports a crash only if it leads to persistent
control-flow hijacking or corrupts state that survives a reset, and a missing
check in an internal helper only if untrusted input reaches it.
When \SystemName scans a pull request, it audits only the changed files, adds
each file's diff, and instructs the Auditor not to report pre-existing bugs in
untouched functions.

\parhead{\texttt{Report Ingestor} agent.}
In \texttt{Re-Analysis} mode, the \texttt{Report Ingestor} agent replaces the
Auditor.
It reads an external report file or directory, such as a Markdown report,
spreadsheet, or log, and emits its findings in the Auditor's schema.
\SystemName loads the resulting findings directly and passes them through the
same downstream stages.

\parhead{\texttt{Deduplicator} agent.}
The \texttt{Deduplicator} agent runs immediately after the entry agent,
so duplicates never consume review or PoC time or tokens.
Within a run, it collapses findings with the same root failure mechanism, such
as several call sites of one unhardened helper, into one canonical finding.
Across runs, it marks a finding as a duplicate if it matches an open finding
from an earlier run of the same project, so repeated scans do not re-report
known issues.
When descriptions are ambiguous, it inspects shared helpers, macros, or structs
with code search tools.

\parhead{\texttt{Adversarial Reviewer} agent.}
The \texttt{Adversarial Reviewer} agent critically reviews the findings of
other agents and rejects those that are unlikely to represent an actual
security-relevant bug (D1, D2).
It reviews each finding twice: the initial review sees only the finding, and
the final review also sees the PoC, its diff, and its harness output.
Its instructions make it skeptical by default.
It cannot rate a finding as exploitable from missing context and must assume
any unlocated definition is implemented safely.
To rate a finding as exploitable or assign high or critical severity, it must
instead construct a step-by-step argument from the actual source code.
It also recalibrates each finding's severity, CVSS vector, and CWE, and has the
largest call budget of any agent.

\subsection{Agent Tools}
\label{sec:design-tools}

\begin{table}[t]
    \centering
    \footnotesize
    \def\y{\ding{51}}
    \setlength{\tabcolsep}{3pt}
    \begin{tabular}{@{}l cccccr@{}}
        \toprule
        \textbf{Agent} & \textbf{File} & \textbf{Search} & \textbf{CWE} &
        \textbf{Expert} & \textbf{Worktree} & \textbf{Calls} \\
        \midrule
        Project Expert        & \y & \y &    &    &    & \callsExpert \\
        Source File Auditor   & \y & \y & \y & \y &    & \callsAuditor \\
        Report Ingestor       & \y &    & \y &    &    & \callsIngestor \\
        Deduplicator          & \y & \y &    & \y &    & \callsDeduplicator \\
        Adversarial Reviewer  & \y & \y & \y & \y &    & \callsReviewer \\
        PoC Exploit Generator & \y & \y & \y & \y & \y & \callsExploiter \\
        \bottomrule
    \end{tabular}
    \caption{Tool categories available to each agent
    (Appendix~\ref{sec:appendix-tools}) and maximum LLM calls per invocation.}
    \label{tab:agent-tools}
\end{table}

General-purpose agents can explore repositories and run build commands, but
reproducing findings in RoT projects requires access to each project's
specialized toolchain and test harness.
\SystemName provides a common execution interface that lets agents invoke these
environments without project-specific agent tools.
Table~\ref{tab:agent-tools} lists each agent's tool categories and LLM call
limit, and Appendix~\ref{sec:appendix-tools} describes each tool.
Beyond standard file access, these tools fall into four categories: code
search, CWE lookup, \texttt{Project Expert} queries, and worktree execution.
No agent has web access, so agents cannot look up known advisories.

For code navigation, \SystemName wraps three external programs in Python tools
that scope each search to the project and limit its output.
These code search tools let agents look up symbol definitions with Universal
Ctags~\cite{universal_ctags_2026}, match structural code patterns on
Tree-sitter~\cite{tree_sitter} syntax trees with ast-grep~\cite{ast_grep_2026},
and search file contents with ripgrep~\cite{ripgrep_2026}.
By chaining these tools, agents trace calls and definitions across files one
step at a time.
A CWE lookup tool searches a bundled copy of MITRE's CWE
catalog~\cite{cwe_mitre} by keyword, and \SystemName validates every finding's
CWE against the same catalog.

Rather than wrapping each project's execution environment in a dedicated tool,
agents run build and test commands (e.g., Cargo, Bazel, and Verilator) through a
single harness command inside the project's own Nix development shell.
As a result, adding a project means adding a new development shell, not a new
tool (D3).
Only the \texttt{PoC Exploit Generator} agent has this harness command, along
with tools that write, patch, diff, and reset files in a scratch worktree.

\subsection{PoC Reproduction and Final Review}
\label{sec:design-verification}

\parhead{\texttt{PoC Exploit Generator} agent.}
The \texttt{PoC Exploit Generator} agent attempts to write a working PoC test
for each finding that survives the initial review, or that enters through the
\texttt{Report Ingestor} agent, attaching executable evidence for maintainers
(D2).
It writes each PoC as a test in the project's own test infrastructure.
Its instructions direct it to inspect the tests and build targets nearest the
affected file first, and to reuse the project's integration models, test
helpers, and mocks rather than build a harness from scratch.
Firmware PoCs run with Cargo for Caliptra (including full-firmware mailbox
tests on Caliptra's hardware model) and Bazel for OpenTitan's ROM\_EXT and
OpenPRoT\@.
RTL PoCs are testbenches that the agent compiles and runs with
Verilator~\cite{verilator_2026}.
The agent works in a scratch Git worktree of the project, where it can write,
patch, diff, and reset files and run build and test commands
(\S\ref{sec:design-tools}).
To bound cost, \SystemName attempts PoCs only for findings at or above a
configurable severity, which defaults to Medium.
Our evaluation sets this threshold to Informational, so every finding that
survives the initial review receives a PoC attempt.

\parhead{Diff check.}
The agent reports whether its PoC verified the finding, and a gate
checks that claim against the worktree.
The gate marks the PoC unverified unless its diff is non-empty, at least one
harness command ran, and the diff changes only test code.
Test code is any file whose path contains a test token, such as
\texttt{test}, \texttt{mock}, \texttt{fuzz}, or \texttt{harness}, along with
build files.
Projects can add their own tokens, and OpenTitan adds \texttt{dv} and
\texttt{tb} for RTL testbenches.
Because Rust places unit tests inside source files, a change to a production
file also counts as test code if it deletes no lines and contains an inline
test marker such as \texttt{\#[cfg(test)]}.
This rule is coarse, and would also accept a production change made alongside
an inline test.

The gate can only downgrade the agent's claim, never upgrade it.
It prevents an agent from manufacturing a proof by weakening a production check
or deleting an assertion.
The gate does not check the test's exit code, because a test may pass either
by triggering the bug or by asserting benign behavior.
Whether the PoC demonstrates the bug is instead left to the final
review.
\SystemName also validates each harness command before running it.
It rejects commands that reference paths outside the worktree, the run's
workspace, the project checkout, the Nix store, and \texttt{/dev}.
It also rejects commands that start with file-exploration programs such as
\texttt{find}, \texttt{grep}, or \texttt{cat}, for which agents must use the
scoped search tools instead.
Each command runs with a sanitized environment and times out after
\harnessTimeout.

\parhead{Final review.}
In the final review, the \texttt{Adversarial Reviewer} agent receives the
finding, the agent's PoC explanation, the actual worktree diff, and the last
harness command with its exit code and output.
It rejects a PoC that modifies production code, mocks the target logic, or
makes only vacuous assertions.
It also rejects a PoC that calls an internal helper directly when the public
entry point already sanitizes the input.
Findings that depend only on debug-build behavior, such as arithmetic overflow
traps, are marked not exploitable.
A failed PoC alone does not close a finding.
If the harness cannot exercise the flaw, for example because it depends on
hardware registers or physical fault injection, the agent labels it a harness
limitation, and the finding stays open as unverified.
The \texttt{Adversarial Reviewer} agent closes a finding with a failed PoC only
if the attempt uncovers a guard preventing exploitation.

\subsection{Input Modes: Discovery vs. Re-Analysis}
\label{sec:design-modes}

\SystemName supports two input modes that share one pipeline.
In \texttt{Discovery} mode, the \texttt{Source File Auditor} agent scans the
project's source files for new security-relevant bugs.
In \texttt{Re-Analysis} mode, the \texttt{Report Ingestor} agent instead reads
an externally generated report or collection of reports.
\SystemName still relies on the project's source files for context, but
focuses on verifying the reported bugs, not identifying new ones.
The same downstream stages process each reported finding, so it leaves the
pipeline with the same verdict, PoC evidence, and recorded reasons as a
discovered one.

\subsection{Hardware Analysis}
\label{sec:design-hardware}

\SystemName analyzes RTL with the same agents, tools, and pipeline as
firmware.
An RTL scan job differs only in its configuration and in the
language-specific guidance that the \texttt{Source File Auditor} agent
receives.
It selects SystemVerilog sources, excludes generated register files and
verification code, and adds test-path tokens for testbenches.
For OpenTitan, the RTL job scans the Earl Grey top level
(\S\ref{sec:bg-rot}) and uses the same threat model as the firmware jobs.
The \texttt{Project Expert} agent's instructions already cover hardware
concerns, including register maps, lock registers such as \texttt{REGWEN},
alert lines, and memory protection, so its summary carries them into every RTL
audit.
Universal Ctags indexes SystemVerilog, letting agents trace module, signal, and
parameter definitions.

RTL findings pass through the same verification as firmware findings.
The \texttt{PoC Exploit Generator} agent writes a standalone C++ or
SystemVerilog testbench in a test path, then compiles and runs it with
Verilator through the same harness command and under the same gate.
A Verilator testbench simulates individual modules, not the full chip.
Findings that a module-level testbench cannot exercise, such as those that
require full-chip simulation or FPGA infrastructure, stay open as
unverified harness limitations (\S\ref{sec:design-verification}).

\section{Implementation}
\label{sec:implementation}

We build \SystemName on Google's Agent Development Kit (ADK)
framework~\cite{adk_2026} for its multi-provider support, tunable system
prompts, and structured input and output schemas.
\SystemName is model-agnostic: besides Gemini, it supports Claude, OpenAI, and
local models via Ollama~\cite{ollama_2026}.
All agents in a run use the same model, and our evaluation uses both Gemini and
Claude models.
We package \SystemName and each project's build environment with
Nix~\cite{nixos_flakes_2026} so that dependency resolution and PoC builds are
reproducible across hosts.
In total, \SystemName comprises \numAgents agents and \numTools tools in
\locPipeline lines of Python for orchestration, \locAgents for agents,
\locTools for tools, \locPrompts lines of prompts, and \locNix of Nix.

\parhead{Orchestration.}
\label{sec:impl-orchestration}
\SystemName builds each run as an ADK workflow graph from an ordered list of
phases, and passes the finding records from one phase to the next.
The Auditor and review phases process their inputs in parallel batches of a
configurable size, with staggered starts to avoid bursts of model requests.
PoC creation instead processes one finding at a time, so PoCs do not contend
for build locks and can reuse the project's build caches.
Every model call retries up to five times with exponential backoff and jitter,
from 2 to 60~s.
Each phase checkpoints its findings to disk, and an interrupted run falls back
to its latest checkpoint.
For Gemini models, \SystemName requests an explicit cache of each phase's system
instructions and tool declarations, with a two-hour lifetime.
For other models, or if the provider rejects the cache request, \SystemName
relies on the provider's built-in prompt caching.
Cache hit rates differ widely across models.
\SystemName also records token usage and tool calls per phase and per agent,
which we use to measure cost.

\parhead{Per-project configuration.}
\label{sec:impl-projects}
Onboarding a project to \SystemName requires four files: a Nix project
definition, a development shell with the project's toolchain, a threat model,
and one or more scan jobs that select the files to analyze (D3).
These files can live in the adopting project's own repository, which makes
integration simpler and does not require the involvement of \SystemName's
maintainers.
Each Caliptra project definition is \cfgCaliptraSwProject lines of Nix, and the
three repositories share one \cfgCaliptraSwShell-line development shell.
OpenTitan's configuration is the largest: its project definition is
\cfgOpentitanProject lines, mostly lists of test files to exclude, and its
\cfgOpentitanShell-line shell configures Bazel and Verilator.
Its \cfgOpentitanNumJobs scan jobs, \cfgOpentitanJobs lines in total, cover both
C firmware and SystemVerilog RTL.
Threat models range from \cfgThreatModelMin to \cfgThreatModelMax lines.
For the evaluation, each target also has a configuration file that pins its
commit and analysis scope.

\parhead{PoC execution.}
\label{sec:impl-poc}
\SystemName runs every PoC in a single scratch Git worktree of the project.
Before each finding, it resets the worktree to the project's commit while
keeping build outputs and the Cargo cache, so consecutive PoCs do not rebuild
the project from scratch.
Agent tools resolve every file path inside the worktree and reject paths that
escape it.
Harness commands run with an allowlist of environment variables, such as
\texttt{PATH}, \texttt{HOME}, and toolchain variables, so that credentials in
\SystemName's own environment do not reach the PoC.
After each finding, \SystemName saves the PoC's diff, a log of every harness
command and its output, and a standalone script that applies the diff and
reruns the last command.
Developers can use this script to reproduce a PoC outside \SystemName.

\begin{figure}[t]
    \centering
    \begin{tikzpicture}[
    box/.style={mj/box, minimum height=6mm},
    trig/.style={box, minimum width=20.33mm},
    prov/.style={box, minimum width=12mm},
    envb/.style={box, minimum width=18mm},
    nixb/.style={box, minimum height=8mm},
    outb/.style={box, minimum width=31.25mm},
    grp/.style={mj/group, inner sep=1.5mm},
    glabel/.style={mj/label, font=\scriptsize\itshape,
      text height=1.5ex, text depth=0.25ex},
  ]
  \def\rowstep{11mm}

  \node[trig] (sj) at (0,0) {Scheduled Job};
  \node[trig, left=1.5mm of sj] (pr) {Pull Request};
  \node[trig, right=1.5mm of sj] (mr) {Manual Run};
  \begin{scope}[on background layer]
    \node[grp, fit=(pr)(mr), label={[glabel]above:Triggered by}] (trig) {};
  \end{scope}

  \path (trig.south) ++(0,-\rowstep) coordinate (r2);
  \node[prov, anchor=north] (cla) at (r2 -| -12.5mm,0) {Claude};
  \node[prov, left=1.5mm of cla] (gem) {Gemini};
  \node[prov, right=1.5mm of cla] (oai) {OpenAI};
  \node[envb, anchor=north] (gh) at (r2 -| 23mm,0) {GitHub CI};
  \begin{scope}[on background layer]
    \node[grp, fit=(gem)(oai), label={[glabel, name=lmodel]above:Model provider}] (model) {};
    \node[grp, fit=(gh), label={[glabel, name=lenv]above:Execution environment}] (envg) {};
  \end{scope}

  \path (model.south) ++(0,-\rowstep) coordinate (r3);
  \node[nixb, minimum width=22mm, anchor=north west] (adk) at (r3 -| -32.3mm,0)
    {\SystemName\\pipeline (ADK)};
  \node[nixb, minimum width=21mm, right=1.5mm of adk] (cfg) {Project\\configuration};
  \node[nixb, minimum width=18mm, right=1.5mm of cfg] (src) {Project\\source code};
  \begin{scope}[on background layer]
    \node[grp, fit=(adk)(src), label={[glabel, name=lnix]above:Nix-managed execution}] (nix) {};
  \end{scope}

  \path (nix.south) ++(0,-\rowstep) coordinate (r4);
  \node[outb, anchor=north east] (gcs) at (r4 -| -0.75mm,0) {Cloud Storage};
  \node[outb, anchor=north west] (dash) at (r4 -| 0.75mm,0) {Generated dashboard};
  \begin{scope}[on background layer]
    \node[grp, fit=(gcs)(dash), label={[glabel, name=lout]above:Output results}] (out) {};
  \end{scope}

  \draw[mj/arr] (trig.south -| lmodel.north) -- (lmodel.north);
  \draw[mj/arr] (trig.south -| lenv.north) -- (lenv.north);
  \draw[mj/arr] (model.south) -- (model.south |- lnix.north);
  \draw[mj/arr] (envg.south) -- (envg.south |- lnix.north);
  \draw[mj/arr] (nix.south) -- (nix.south |- lout.north);
\end{tikzpicture}
    \caption{\SystemName's deployment architecture.}
    \label{fig:deployment}
\end{figure}
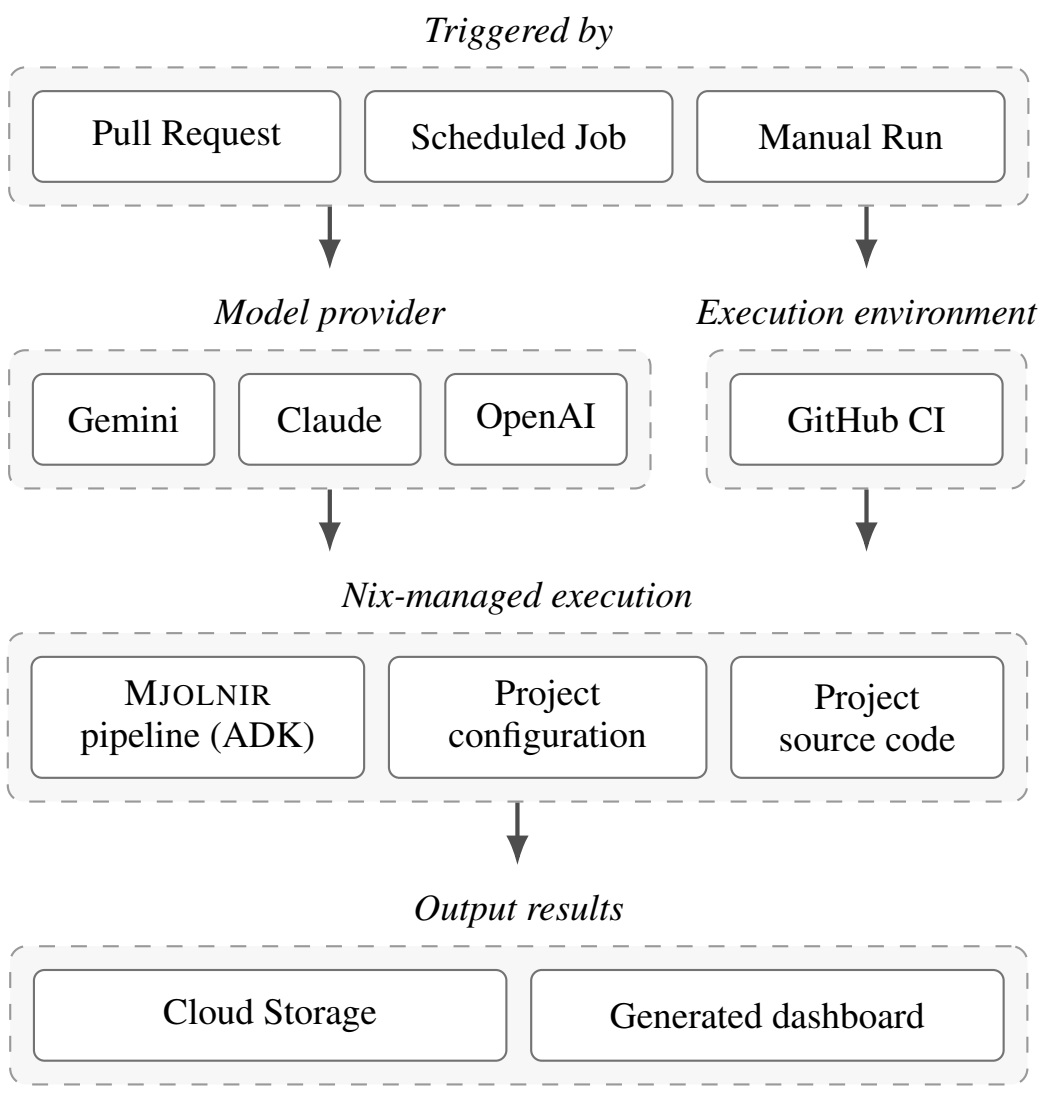

\parhead{Deployment.}
\label{sec:impl-deployment}
Figure~\ref{fig:deployment} shows how \SystemName is deployed to real-world RoT
projects.
An adopting project runs \SystemName from its own CI on pull requests, on a
schedule, or manually.
On a pull request, \SystemName runs over the changed files and can post its
findings directly to the pull request as review comments.
Each run writes its findings, logs, and token and tool usage to a run directory
that \SystemName syncs to cloud storage.
The \texttt{Deduplicator} agent reads open findings from earlier runs in this
storage, so repeated scans do not re-report them.
A static web dashboard built from the stored runs shows each finding's verdict,
CVSS score, history, duplicates, PoC status, and reproducer, along with the flow
of findings through each phase.
Three production RoT repositories have deployed this CI workflow, auditing over
800 pull requests over four months with maintainers inspecting findings in the
external web dashboard during code review.

\section{Evaluation}
\label{sec:evaluation}

Our evaluation answers five research questions:
\begin{enumerate}[label=\textbf{RQ\arabic*}, leftmargin=*]
    \item \textbf{Finding quality.}
        How precise are \SystemName's findings on Caliptra's runtime firmware, compared
        to four single-agent baselines and two ablations?

    \item \textbf{Verification funnel.}
        How many findings does each stage close, and how many surviving
        findings receive a verified PoC?

    \item \textbf{Stability.}
        How consistent are \SystemName's findings across repeated scans and
        across models?

    \item \textbf{Known bugs.}
        How many of 34 known security-relevant bugs in Caliptra's core firmware
        does \SystemName find, compared to the best of four single-agent
        baselines, and does any stage close them?

    \item \textbf{Generality.}
        Does \SystemName find new security-relevant bugs in other RoT projects,
        languages, and RTL?
\end{enumerate}

We evaluate \SystemName in three parts: an in-depth study of Caliptra's runtime
firmware (RQ1--RQ3), a known-bug study on Caliptra's core firmware (RQ4), and
scans of five additional targets across Caliptra, OpenTitan, and OpenPRoT (RQ5).
We then report what these scans cost.

\subsection{Experimental Setup}
\label{sec:eval-setup}
\begin{table}[t]
    \footnotesize
    \centering
    \setlength{\tabcolsep}{3.5pt}
    \caption{Evaluation targets for \SystemName.}
    \label{tab:targets}
    \begin{tabular}{@{}lllrr@{}}
        \toprule
        Target & Language & Build \& test & Files & Code \\
        \midrule
        \multicolumn{5}{@{}l}{\textit{Caliptra}} \\
        \quad Runtime firmware & Rust & Cargo & \tgtCaliptraSwFiles & \tgtCaliptraSwCode \\
        \quad MCU ROM, protocols & Rust & Cargo & \tgtCaliptraMcuSwFiles & \tgtCaliptraMcuSwCode \\
        \quad DPE & Rust & Cargo & \tgtCaliptraDpeFiles & \tgtCaliptraDpeCode \\
        \multicolumn{5}{@{}l}{\textit{OpenTitan}} \\
        \quad ROM\_EXT firmware & C & Bazel & \tgtOpentitanFwFiles & \tgtOpentitanFwCode \\
        \quad Earl Grey RTL & SystemVerilog & Verilator & \tgtOpentitanRtlFiles & \tgtOpentitanRtlCode \\
        \multicolumn{5}{@{}l}{\textit{OpenPRoT}} \\
        \quad Platform services & Rust & Bazel, Pigweed & \tgtOpenprotFiles & \tgtOpenprotCode \\
        \midrule
        \textbf{Total} & & & \tgtTotalFiles & \tgtTotalCode \\
        \midrule
        \multicolumn{5}{@{}l}{\textit{Known-bug study (RQ4)}} \\
        \quad Caliptra core firmware & Rust & Cargo & \tgtCaliptraCoreFiles & \tgtCaliptraCoreCode \\
        \bottomrule
    \end{tabular}
\end{table}

\parhead{Targets and isolation.}
Table~\ref{tab:targets} lists our six evaluation targets, which together contain
\tgtTotalFiles files and \tgtTotalCode lines of code, and the Caliptra core
firmware release that we scan for known bugs.
We scan each target at a fixed commit, in a fresh workspace that contains only
that commit, with no Git history or remote branches available.
Each trial runs with the outputs of other trials and \SystemName's caches
hidden, so trials cannot see each other's findings.
Agents have no web tools.
For RQ1--RQ3 and RQ5, we count only previously unknown security-relevant bugs,
whose fixes and advisories cannot appear in model training data.
For RQ4, we scan a commit with known bugs fixed after the models' training
cutoffs.

\parhead{Configurations.}
On every target, we run \SystemName with the PoC severity threshold set to
Informational, so every finding that survives the initial review process
receives a PoC attempt.
From each run, we extract five checkpoints from the finding records' history:
after discovery, deduplication, initial review, PoC creation, and final review
(\S\ref{sec:design-pipeline}).
On Caliptra's runtime firmware, we also run two ablations of \SystemName: one
without the threat model and one without the \texttt{Project Expert} agent.

\parhead{Single-agent baseline.}
Our performance baseline is a single ADK agent that uses the same model as
\SystemName but none of its pipeline or prompts.
It has no \texttt{Project Expert} agent, deduplication, review, or PoC
execution.
Its instructions are a generic security-audit prompt that asks for concrete,
exploitable vulnerabilities, excludes theoretical issues that existing checks
prevent, and requests the same report fields as \SystemName's Auditor, with no
RoT-specific guidance.
We run it in four configurations, on Caliptra's runtime firmware and in the
known-bug study: two modes, each with and without the threat model.
In whole-directory mode, one session audits the source directory with only
file-listing, file-reading, and text-search tools.
In per-file mode, each file gets its own session with no tools, and the agent
receives the file's contents in its prompt.
Because the baseline shares \SystemName's model, differences between them
reflect \SystemName's prompts and pipeline rather than a stronger model.

\parhead{Models and repetitions.}
All agents in a run use the same model, which defaults to Gemini~3.8 Flash
for its availability and low list price.\footnote{In our evaluation, Gemini~4
Argon was nonetheless cheaper per run than Gemini~3.8 Flash because it read
fewer tokens (Appendix~\ref{sec:appendix-cost}).}
Both Gemini~3.8 Flash and Gemini~4 Argon have a March 2026 training
cutoff~\cite{google_gemini_3_8_flash_2026,google_gemini_4_argon_2026}, and
Claude Sonnet~5.5 has a June 2026 cutoff~\cite{anthropic_training_data_2026}.
On Caliptra's runtime firmware, we run \SystemName, each ablation, and
each single-agent baseline three times with Gemini~3.8 Flash.
We also run \SystemName three times each with Gemini~4 Argon and
Claude Sonnet~5.5 to measure consistency across models.
To measure consistency within one model at greater depth, we extend the
three \SystemName runs on Caliptra's runtime firmware with two
additional repetitions for a five-repetition stability study.
On the remaining five targets, we run \SystemName once with Gemini~3.8 Flash to
evaluate generality across projects, languages, and build harnesses.
In the known-bug study, we run \SystemName and each single-agent configuration
once with Gemini~3.8 Flash and once with Gemini~4 Argon.
For configurations with more than one run, we report per-run counts and costs as
means across runs, whereas expert labels, cross-run issues, and safeguard events
are aggregated across runs.

\begin{table*}[t]
    \footnotesize
    \centering
    \setlength{\tabcolsep}{4pt}
    \caption[Results per configuration on Caliptra's runtime firmware over three runs.]{Results per configuration on Caliptra's runtime firmware over three
    runs. \textbf{Open}, \textbf{PoC}, and \textbf{Cost} are per-run means
    ($\pm$SD). \textbf{TP}, \textbf{FP}, and \textbf{Precision} count
    expert-labeled findings across the three runs: all open findings, or a
    random sample of 4 per run ($N=12$) for rows marked~$^\dagger$.}
    \label{tab:rt-results}
    \begin{tabular}{@{}lrrrrrr@{}}
        \toprule
        & & \multicolumn{3}{c}{\textbf{Expert-Labeled (3 runs)}} & & \\
        \cmidrule(lr){3-5}
        \textbf{Configuration} & \textbf{Open} & \textbf{TP} & \textbf{FP} & \textbf{Precision} & \textbf{PoC} & \textbf{Cost} \\
        \midrule
        \SystemName & \evalRtOpenMeanSd & \evalRtTp & \evalRtFp & \evalRtPrecision & \evalRtPocVerifiedMeanSd & \evalRtCost \\
        \texttt{w/o Threat Model} & \evalRtNoTmOpenMeanSd & \evalRtNoTmTp & \evalRtNoTmFp & \evalRtNoTmPrecision & \evalRtNoTmPocVerifiedMeanSd & \evalRtNoTmCost \\
        \texttt{w/o Project Expert} & \evalRtNoExpertOpenMeanSd & \evalRtNoExpertTp & \evalRtNoExpertFp & \evalRtNoExpertPrecision & \evalRtNoExpertPocVerifiedMeanSd & \evalRtNoExpertCost \\
        Single-Agent (Per-File, w/o TM)$^\dagger$ & \evalRtPerFileNoTmOpenMeanSd & \evalRtPerFileNoTmTp & \evalRtPerFileNoTmFp & \evalRtPerFileNoTmPrecision & -- & \evalRtPerFileNoTmCost \\
        Single-Agent (Per-File)$^\dagger$ & \evalRtPerFileOpenMeanSd & \evalRtPerFileTp & \evalRtPerFileFp & \evalRtPerFilePrecision & -- & \evalRtPerFileCost \\
        Single-Agent (Whole-Dir, w/o TM) & \evalRtWholeDirNoTmOpenMeanSd & \evalRtWholeDirNoTmTp & \evalRtWholeDirNoTmFp & \evalRtWholeDirNoTmPrecision & -- & \evalRtWholeDirNoTmCost \\
        Single-Agent (Whole-Dir) & \evalRtWholeDirOpenMeanSd & \evalRtWholeDirTp & \evalRtWholeDirFp & \evalRtWholeDirPrecision & -- & \evalRtWholeDirCost \\
        \SystemName (Gemini 4 Argon)$^\dagger$ & \evalRtArgonOpenMeanSd & \evalRtArgonTp & \evalRtArgonFp & \evalRtArgonPrecision & \evalRtArgonPocVerifiedMeanSd & \evalRtArgonCost \\
        \SystemName (Claude Sonnet 5.5) & \evalRtSonnetOpenMeanSd & \evalRtSonnetTp & \evalRtSonnetFp & \evalRtSonnetPrecision & \evalRtSonnetPocVerifiedMeanSd & \evalRtSonnetCost \\
        \bottomrule
    \end{tabular}
\end{table*}

\parhead{Labeling and metrics.}
Because unknown bugs have no ground truth, we measure recall only on the known
bugs in RQ4 and evaluate open findings by precision, $\text{TP} / (\text{TP} +
\text{FP})$, over expert labels.
Across the 27 runs (nine configurations) on Caliptra's runtime firmware, 348
findings remain open, and three Caliptra maintainers from two organizations
blindly label \evalCaliptraSwPoolInSample of them: all 102 from the six
configurations averaging at most ten open findings per run, plus four randomly
sampled per run (36 total) from the three averaging over twenty.
Reviewers mark a finding as a true positive (TP) if it is a security-relevant
bug (\S\ref{sec:threat-model}) and as a false positive (FP) otherwise.
They work blind to configuration, stage, and model, with each reviewer's
findings grouped by source file.
We count each maintainer-confirmed open finding of \SystemName as a confirmed
finding rather than deduplicating bugs across runs, and we treat verified PoCs
as evidence rather than labels.
Project maintainers label open findings on the other five targets under the same
definition.
Within each multi-agent run, stable per-finding identifiers track stage closures
across checkpoints without labels.

The primary Caliptra reviewer labels all \evalCaliptraSwPoolInSample sampled
findings, a second labels \evalCaliptraSwPoolDoubleReviewed, and a third labels
9 of those \evalCaliptraSwPoolDoubleReviewed.
On the \evalCaliptraSwPoolDoubleReviewed multiply labeled findings, raw
pairwise agreement is 51\% and Krippendorff's
$\alpha=-0.04$~\cite{krippendorff_content_2018}: the second reviewer weighed
standalone exploitability more heavily than missing countermeasures, marking 32
of 56 as false positives versus 15 for the primary reviewer, and 13 of the 26
disputed findings recur at two code locations across runs.
We adjudicate all 26 against the pinned source code and threat model, resolving
20 as true positives and 6 as false positives (5 where the dissenting reviewer
spotted an overlooked guard).
If we exclude all 26 instead, \SystemName remains at \evalRtPrecision precision
(18 of 18) and the best baseline rises from \evalRtWholeDirPrecision to 81.8\%.

\subsection{Caliptra Runtime Firmware}
\label{sec:eval-runtime}

RQ1--RQ3 study \SystemName in depth on Caliptra's runtime firmware, where we
compare it to single-agent baselines, ablations, and other models.

\subsubsection{RQ1: Finding Quality}
\label{sec:eval-quality}

Table~\ref{tab:rt-results} compares open findings per run, labeled true and
false positives, precision, PoC-verified findings, and cost across
configurations on Caliptra's runtime firmware.
Across the three models, maintainers confirm \evalConfirmedRt of \SystemName's
labeled findings as true positives: all \evalRtTp open findings from the three
Flash runs (18 directly from maintainer labels and 6 after adjudication, for
\evalRtPrecision precision), \evalRtArgonTp of \evalRtArgonSampled sampled with
Gemini~4 Argon (\evalRtArgonPrecision), and all \evalRtSonnetTp with Claude
Sonnet~5.5 (\evalRtSonnetPrecision).
The single-agent baselines fail in two different ways: per-file auditing
reports the most findings, but one in four is a false positive because an agent
that sees one file at a time misses checks enforced in callers and flags
non-security checks, whereas whole-directory auditing is more precise
(\evalRtWholeDirPrecision) but reports the fewest findings
(\evalRtWholeDirOpenMeanSd per run).
Removing the threat model lowers precision for the whole-directory baseline
from \evalRtWholeDirPrecision to \evalRtWholeDirNoTmPrecision (while the
per-file baseline stays at \evalRtPerFilePrecision on its 12 sampled findings),
and removing either the threat model or the \texttt{Project Expert} agent
lowers \SystemName's precision to \evalRtNoTmPrecision and
\evalRtNoExpertPrecision.
In both ablations, the false positives are findings the \texttt{Adversarial
Reviewer} agent rejects when it has project context, such as documented
trust-boundary conventions reported as authentication bypasses.

Overall, project context raises precision on the same model from
\evalRtWholeDirNoTmPrecision--\evalRtWholeDirPrecision for the single-agent
baselines to \evalRtPrecision for \SystemName, and removing either component of
that context reintroduces false positives.

\subsubsection{RQ2: Verification Funnel}
\label{sec:eval-funnel}

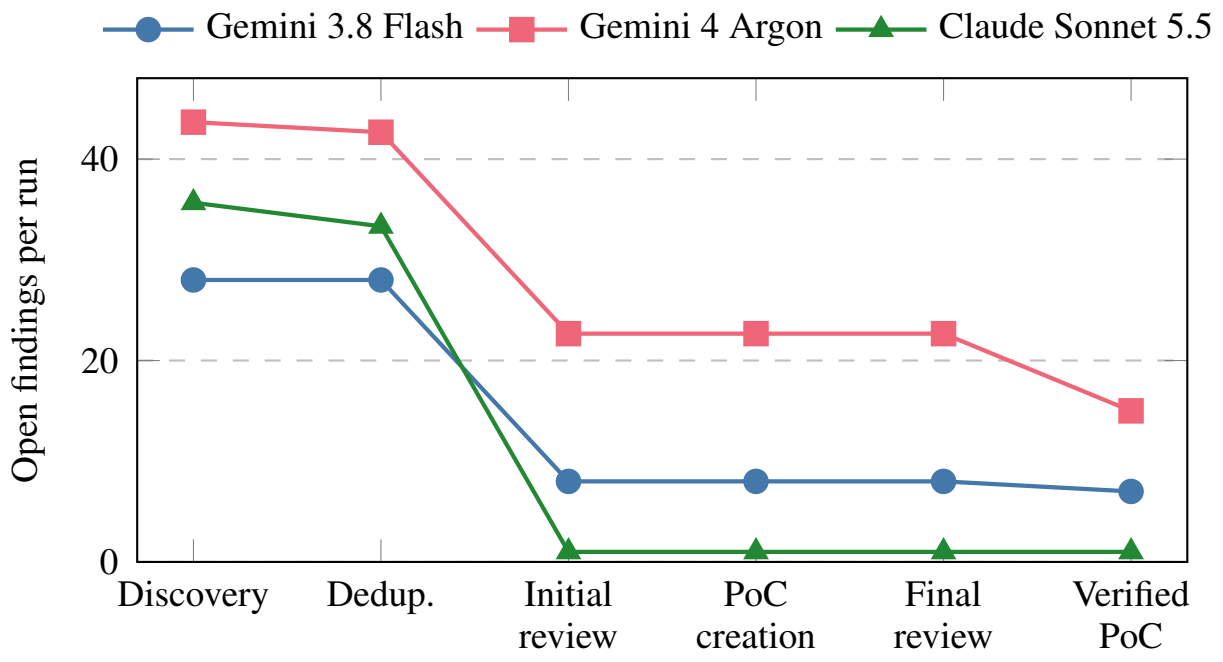
\begin{figure}[t]
    \centering
    \begin{tikzpicture}
\begin{axis}[
    scale only axis, width=0.82\columnwidth, height=3.2cm,
    symbolic x coords={discovery,deduplication,initial_review,poc_creation,final_review,poc}, xtick=data, xticklabels={{Discovery},{Dedup.},{Initial\\review},{PoC\\creation},{Final\\review},{Verified\\PoC}},
    enlarge x limits=0.06, xticklabel style={align=center},
    ymin=0, ymajorgrids=true, grid style=dashed,
    ylabel={Open findings per run},
    xticklabel style={font=\scriptsize}, yticklabel style={font=\scriptsize},
    ylabel style={font=\scriptsize},
    legend style={at={(0.5,1.02)}, anchor=south, legend columns=3, font=\scriptsize, draw=none},
]
\addplot[color={rgb,255:red,68;green,119;blue,170}, mark=*, thick] coordinates { (discovery,28.00) (deduplication,28.00) (initial_review,8.00) (poc_creation,8.00) (final_review,8.00) (poc,7.00) };
\addlegendentry{Gemini 3.8 Flash}
\addplot[color={rgb,255:red,238;green,102;blue,119}, mark=square*, thick] coordinates { (discovery,43.67) (deduplication,42.67) (initial_review,22.67) (poc_creation,22.67) (final_review,22.67) (poc,15.00) };
\addlegendentry{Gemini 4 Argon}
\addplot[color={rgb,255:red,34;green,136;blue,51}, mark=triangle*, thick] coordinates { (discovery,35.67) (deduplication,33.33) (initial_review,1.00) (poc_creation,1.00) (final_review,1.00) (poc,1.00) };
\addlegendentry{Claude Sonnet 5.5}
\end{axis}
\end{tikzpicture}
    \caption{Mean open findings per run after each stage of \SystemName on
    Caliptra's runtime firmware.}
    \label{fig:funnel}
\end{figure}

RQ2 follows each \SystemName run through its five checkpoints without expert
labels (Figure~\ref{fig:funnel}).

The initial review is the only stage that closes many findings, removing
\evalRtInitRevDropPct of findings entering it on Gemini~3.8 Flash,
\evalRtArgonInitRevDropPct on Gemini~4 Argon, and
\evalRtSonnetInitRevDropPct on Claude Sonnet~5.5.
Deduplication removes few findings because each run starts with an empty
history (\S\ref{sec:eval-setup}), and PoC creation and the final review attach
executable evidence rather than filter: most surviving findings receive a
verified PoC (Table~\ref{tab:rt-results}).
Findings that stay open without a verified PoC are mostly ones the
\texttt{Adversarial Reviewer} agent keeps open as harness limitations, such as
\evalRtArgonHarnessLimited of Gemini~4 Argon's \evalRtArgonOpenUnverified
unverified findings per run.
Closing a finding does not prove it false, and RQ4 measures how many known bugs
the review closes.

\subsubsection{RQ3: Stability Across Scans and Models}
\label{sec:eval-stability}

RQ3 examines how stable \SystemName's output is across repeated runs of one
model and across models.
To check whether runs report the same bugs, not only similar counts
(Table~\ref{tab:rt-results} and Figure~\ref{fig:funnel}), we match open
findings across runs without expert labels: two findings match if they are in
the same file and their line spans are within \overlapTolerance lines of each
other (or share a CWE when no line span is given), forming one issue.
This matching is approximate and can merge distinct bugs in one function or
split one bug located differently.

Extending \SystemName's three Gemini~3.8 Flash runs with two additional
repetitions yields \evalRtStabilityDiscOpen candidate findings at discovery,
\evalRtStabilityOpenMeanSd open findings after final review, and
\evalRtStabilityPocVerifiedMeanSd PoC-verified findings per run.
Across the five runs, \SystemName reports \overlapStabIssues distinct issues:
\overlapStabMulti recur across multiple runs (\overlapStabMajority in at least
three, including \overlapStabAllRuns in every run, and \overlapStabTwo in two
runs), while \overlapStabSingle appear in only one (\overlapStabRecurringPct
of a run's issues recur in another run on average).
Across models, all three propose similar numbers of candidate findings at
discovery (Figure~\ref{fig:funnel}), but diverge during the initial review,
where Sonnet's agents invoke tools far less often than either Gemini model
and filter aggressively.
Across their three runs each, Gemini~4 Argon, Gemini~3.8 Flash, and Claude
Sonnet~5.5 leave \overlapModelArgonIssues, \overlapModelFlashIssues, and
\overlapModelSonnetIssues distinct issues open (\overlapModelIssues in total):
\overlapModelAll of Sonnet's \overlapModelSonnetIssues issues are reported by
all three models, \overlapModelTwo more are shared by two models, and
\overlapModelOne appear in only one model (mostly Argon's broader retained
set).
A single run therefore covers only part of what \SystemName finds, and repeated
scans or scans with several models find more.

\subsection{RQ4: Known Bugs}
\label{sec:eval-known-bugs}

RQ4 asks how many known security-relevant bugs \SystemName finds, compared to a
single-agent baseline.
We scan Caliptra's core firmware at release tag \texttt{fw-2.1.0} (commit
\texttt{7da3b01}, March 2026), covering all ten on-target firmware crates,
including the ROM, first mutable code (FMC), runtime, and shared drivers
(\tgtCaliptraCoreFiles files and \tgtCaliptraCoreCode lines of code).
To build a list of known bugs, we audit all 295 upstream commits merged after
\texttt{fw-2.1.0} through October 27, 2026, and keep each fix that was merged
after both Gemini models' March 2026 training
cutoff~\cite{google_gemini_3_8_flash_2026,google_gemini_4_argon_2026} and
addresses a CWE-classified security defect whose defective code is already
present at \texttt{fw-2.1.0} in these crates.
This yields 34 known security-relevant bugs in three tiers: 5 with
published CVEs, 22 security bugs that maintainers fixed without a CVE, and 7
zeroization, attestation, and side-channel hardening fixes
(Appendix~\ref{sec:appendix-known-bugs}).
Only Tier~A carries upstream security advisories.
One Tier~A bug, CVE-2026-7328, was originally found by an early \SystemName
prototype that predates the pipeline described and evaluated here, none of which
it used.

We identify and classify Tiers~B and~C from the commit log
using our definition of a security-relevant bug (\S\ref{sec:threat-model}),
which also covers in-development features and defense-in-depth checks that do
not warrant an advisory.
We run \SystemName and the four single-agent configurations once each with
Gemini~3.8 Flash and once each with Gemini~4 Argon (omitting Claude Sonnet~5.5,
whose June 2026 cutoff postdates early fixes), in the same isolated workspaces
as our other scans (\S\ref{sec:eval-setup}).
We compare each \SystemName run with the best single-agent configuration that
uses the same model, rather than with the union of all four.
A run finds a known bug if one of its open findings is in a file that contains
the bug, overlaps the bug's defective lines to within 30 lines or names one of
its functions, and mentions at least one term from each of the bug's keyword
groups.
Because every finding keeps its identifier across stages, matching the
findings of each checkpoint in the same way also shows whether a known bug
found at discovery is later closed, and by which stage
(Table~\ref{tab:known-bugs}, bottom).
The single-agent baseline has no such stages.

\begin{table}[t]
    \footnotesize
    \centering
    \setlength{\tabcolsep}{3pt}
    \caption{Known bugs found in Caliptra core firmware.}
    \label{tab:known-bugs}
    \begin{tabular}{@{}lrrrrr@{}}
        \toprule
        & & \multicolumn{2}{c}{Gemini 3.8 Flash} & \multicolumn{2}{c@{}}{Gemini 4 Argon} \\
        \cmidrule(lr){3-4} \cmidrule(l){5-6}
        Tier & Bugs & \SystemName & Baseline & \SystemName & Baseline \\
        \midrule
        A: CVEs & 5 & \evalKbFlashA & \evalKbFlashBaseA & \evalKbArgonA & \evalKbArgonBaseA \\
        B: Maintainer fixes & 22 & \evalKbFlashB & \evalKbFlashBaseB & \evalKbArgonB & \evalKbArgonBaseB \\
        C: Hardening & 7 & \evalKbFlashC & \evalKbFlashBaseC & \evalKbArgonC & \evalKbArgonBaseC \\
        \midrule
        \textbf{All} & 34 & \evalKbFlashAll & \evalKbFlashBaseAll & \evalKbArgonAll & \evalKbArgonBaseAll \\
        \midrule
        \multicolumn{6}{@{}l}{\SystemName, \textit{all tiers}} \\
        \quad Found at discovery & & \evalKbFlashDisc & -- & \evalKbArgonDisc & -- \\
        \quad Closed by a review & & \evalKbFlashClosed & -- & \evalKbArgonClosed & -- \\
        \quad With verified PoC & & \evalKbFlashPoc & -- & \evalKbArgonPoc & -- \\
        \bottomrule
    \end{tabular}
\end{table}

Table~\ref{tab:known-bugs} shows that \SystemName outperforms every single-agent
baseline on both models, whereas the highest-recall baseline shifts from
\texttt{Per-File} on Flash (\evalKbFlashBaseAll bugs) to \texttt{Whole-Dir, w/o
TM} on Argon (\evalKbArgonBaseAll).
Deduplication closes no known bugs on either model.
With Argon, the review stages retain all \evalKbArgonAll discovered Tier~A and
Tier~B bugs and close only two Tier~C stack-zeroization items because Caliptra's
single-address-space M-mode firmware exposes no standalone stack-read primitive.
With Flash, by contrast, initial review closes all three discovered Tier~C items
and four Tier~B bugs, and final review closes \texttt{CVE-2026-11835}.
The ten Tier~B bugs missed by both \SystemName runs are mainly external
specification checks or cross-reset hardware invariants, such as FIPS~140-3 XTS
key-half inequality, ML-DSA-87 signature padding, and Key Vault write-locks
across resets.

\parhead{Training data contamination.}
All 34 fixes were merged in April 2026 or later, and the five CVEs were
disclosed between June and August 2026, after both Gemini models' March 2026
cutoff.
Because published cutoffs are approximate, we also evaluate each \SystemName run
against same-model baselines under identical offline conditions, holding any
residual pretraining exposure constant.

\subsection{RQ5: Generality}
\label{sec:eval-generality}
\label{sec:findings}

RQ5 asks whether \SystemName finds new security-relevant bugs beyond Caliptra's
runtime firmware.
Table~\ref{tab:generality} reports, for each of the other five targets, the
findings that remain open after the final review, how many have a verified PoC,
how project maintainers classified them (TP vs.\ FP), and cost.

\begin{table}[t]
    \footnotesize
    \centering
    \setlength{\tabcolsep}{3pt}
    \caption{Results on the other five targets (Gemini~3.8 Flash).}
    \label{tab:generality}
    \begin{tabular}{@{}lrrrrr@{}}
        \toprule
        & & & \multicolumn{2}{c}{Maintainer} & \\
        \cmidrule(lr){4-5}
        Target & Open & PoC & TP & FP & Cost \\
        \midrule
        \multicolumn{6}{@{}l}{\textit{Caliptra}} \\
        \quad MCU ROM, protocols & \evalCaliptraMcuSwFinalOpen & \evalCaliptraMcuSwPocVerified & 9 & 1 & \evalCaliptraMcuSwCost \\
        \quad DPE & \evalCaliptraDpeFinalOpen & \evalCaliptraDpePocVerified & -- & -- & \evalCaliptraDpeCost \\
        \multicolumn{6}{@{}l}{\textit{OpenTitan}} \\
        \quad ROM\_EXT firmware & \evalOpentitanFwFinalOpen & \evalOpentitanFwPocVerified & 3 & 0 & \evalOpentitanFwCost \\
        \quad Earl Grey RTL & \evalOpentitanRtlFinalOpen & \evalOpentitanRtlPocVerified & 5 & 0 & \evalOpentitanRtlCost \\
        \multicolumn{6}{@{}l}{\textit{OpenPRoT}} \\
        \quad Platform services & \evalOpenprotFinalOpen & \evalOpenprotPocVerified & 10 & 1 & \evalOpenprotCost \\
        \bottomrule
    \end{tabular}
\end{table}

The funnel behavior seen on Caliptra's runtime firmware (\S\ref{sec:eval-funnel})
holds across the other five targets despite differences in language and build
harness.
On each target, the initial review discards most candidate alerts, closing all
\evalCaliptraDpeDiscOpen on Caliptra's DPE, while \SystemName attaches passing
PoCs across Cargo, Bazel, Pigweed, and Verilator testbenches without changing
its pipeline.
Across OpenTitan and OpenPRoT, maintainers confirmed all 15 PoC-verified
findings as true positives, whereas the sole false positive came from one of the
four open findings without a PoC.
The other three unverified findings were still confirmed as true positives whose
paths the host harnesses could not exercise: one on OpenTitan's ROM\_EXT
firmware and two of the \evalOpenprotHarnessLimited unverified findings on
OpenPRoT.
On Caliptra's MCU firmware, attempting a PoC also uncovered a guard in the code
that static review had missed, allowing the final review to close
\evalCaliptraMcuSwDisprovedByHarness false alert.
Two maintainers labeled all 10 remaining open findings, agreeing on 8 of the 10
(Krippendorff's $\alpha=0.41$), with 9 true positives after adjudicating the
two disagreements against the source code (\S\ref{sec:eval-setup}).
Both maintainers rejected the remaining alert because hardware RTL prevents the
access exercised against the host-side mock.
Across all six targets, maintainers have confirmed \evalConfirmedTotal of
\SystemName's labeled findings as previously unknown security-relevant bugs:
\evalConfirmedRt across the three models' runs on Caliptra's runtime firmware
(\S\ref{sec:eval-quality}, without cross-run deduplication) and
\evalConfirmedOther across the single-run scans of the other five targets
(Table~\ref{tab:generality}).
We withhold identifying details across targets pending upstream fixes.

Overall, without changes to its prompts or pipeline, \SystemName transfers
across C and Rust firmware and SystemVerilog RTL, reproduces findings across
four build and simulation environments, and uncovers previously unknown
security-relevant bugs.

\subsection{Cost}
\label{sec:eval-cost}

A \SystemName run on Caliptra's runtime firmware costs between
\evalRtSonnetCost and \evalRtCost, depending on the model, and between
\evalCaliptraDpeCost and \evalOpentitanRtlCost with Gemini~3.8 Flash on the
other five targets (Tables~\ref{tab:rt-results} and~\ref{tab:generality}).
The single-agent baselines cost \evalRtWholeDirCost--\evalRtPerFileCost per
run, more than an order of magnitude less than \SystemName on the same model,
but run no review and produce no PoCs.
Discovery accounts for the largest share of cost, and
Appendix~\ref{sec:appendix-cost} breaks cost down by token type, stage, and
model, showing that per-token price is a poor guide to cost per run.

\section{Discussion}
\label{sec:discussion}

\subsection{Limitations and Threats to Validity}
\label{sec:disc-limitations}

\parhead{Execution environment.}
Firmware PoCs run as host-side tests against each project's own mocks or, for
Caliptra, its hardware model, and RTL PoCs run as module-level Verilator
testbenches~\cite{verilator_2026}.
Neither replaces the hardware that the firmware runs on in production.
Findings that depend on hardware registers, fault injection, or full-chip
behavior therefore remain unverified even when they are real, and a missing
fault-injection countermeasure can never receive a verified PoC.
FPGA execution is one path toward higher-fidelity PoCs.

\parhead{Diff check.}
The diff check verifies that a PoC ran and changed only test code, not that it
demonstrates the bug.
Its rule for inline Rust tests would accept a production change made alongside
an inline test, and because build files count as test code, a PoC could also
change a build flag that disables a production check.
The final review guards against such PoCs, but because we do not audit every
PoC manually, we treat a verified PoC as evidence for maintainers, not proof.

\parhead{Per-file auditing.}
Each \texttt{Source File Auditor} agent audits one file, and learns about
related code through the threat model, its tools, and the \texttt{Project
Expert} agent.
Multi-file bugs therefore depend on agents finding the related code themselves,
which auditing file clusters could address.

\parhead{Evaluation scope.}
Our targets contain an unknown number of security-relevant bugs, and publicly
known bugs may appear in model training
data~\cite{zhou_lessleak-bench_2025,roberts_cutoff_2024}.
We therefore measure recall only on bugs fixed after the models' training
cutoffs (RQ4), and precision on a sample of open findings labeled by experts.
Because experts label only open findings, we cannot measure how many real bugs
the review closes on Caliptra's runtime firmware.
Our baselines run only on Caliptra's runtime firmware and in the known-bug
study on its core firmware, and our ablations and Claude Sonnet~5.5 runs cover
only its runtime firmware.
The known-bug study and the other five targets run once per configuration, so
small recall differences may reflect run-to-run variation
(\S\ref{sec:eval-stability}).

\subsection{Future Research Directions}
\label{sec:disc-future}

\parhead{Other RoT ecosystems.}
\SystemName is not designed around Caliptra, OpenTitan, or OpenPRoT, and in
principle applies to any RoT project with firmware or RTL\@.
Other ecosystems to explore include Oxide's Hubris-based
firmware~\cite{hubris_2026} and Microsoft's Project
Cerberus~\cite{cerberus_2026}.

\parhead{Learning across scans.}
Today, agents share knowledge within a run through the \texttt{Project Expert}
agent, and across runs only through the \texttt{Deduplicator} agent's record of
past findings.
A final, self-reflective stage could let agents propose changes to a project's
threat model and to their own prompts, so that scans of one codebase improve
over time even with fixed models.
Co-RedTeam keeps long-term memory of vulnerability patterns across
tasks~\cite{he_co-redteam_2026}, whereas we would evolve one project's threat
model over repeated scans.

\parhead{Synthetic vulnerability benchmark.}
A pipeline similar to \SystemName could inject realistic bugs into local copies
of RoT projects, as AVIATOR~\cite{aviator_2026} does for generic software.
Injected bugs would be both absent from training data and fully known.
However, while models rarely refused discovery or PoC tasks in our runs
(Appendix~\ref{sec:appendix-safeguards}), they often refused to synthesize new
vulnerabilities in our preliminary experiments.
Less restricted or fine-tuned local models could make such a benchmark
practical.

\parhead{Exploit understandability.}
\SystemName optimizes PoCs to run, not to be read.
Optimizing PoCs for readability and conciseness would make findings easier for
maintainers to triage.

\section{Related Work}
\label{sec:related}

\parhead{Agentic vulnerability discovery.}
LLM agents now find real vulnerabilities in general-purpose software.
Big Sleep found an exploitable memory-safety bug, a stack buffer underflow, in
SQLite~\cite{big_sleep_team_naptime_2024}, and the cyber reasoning systems in
DARPA's AIxCC discover vulnerabilities in open-source software, prove them with
proofs of vulnerability, and patch them~\cite{zhang_sok_2026,aixcc_2026}.
Other systems combine LLMs with program analysis to reduce false positives.
IRIS has an LLM label taint sources and sinks for a static
analyzer~\cite{li_iris_2024}, LLMSAN validates the data-flow paths that an LLM
reports~\cite{wang_sanitizing_2024}, and Vulnhuntr traces call chains from
remote input in Python code~\cite{vulnhuntr_2026}.
Chai discovers cryptographic misuse and validates it through differential
testing across libraries~\cite{villa_chai_2026}.

The closest systems also verify findings by execution.
OpenAnt decomposes a codebase into units reachable from external input,
verifies each finding adversarially, and runs PoCs in Docker
containers~\cite{korda_openant_2026}.
Co-RedTeam pairs discovery with execution-grounded exploitation and long-term
memory, and its ablations find execution feedback
critical~\cite{he_co-redteam_2026}.
\SystemName shares this pipeline shape, which we do not claim as novel, but
neither system reaches RoT code.
Co-RedTeam is evaluated only on benchmarks, and OpenAnt analyzes only code
reachable from HTTP, WebSocket, command-line, or file inputs, none of which our
targets expose.

\parhead{PoC generation and security benchmarks.}
PAGENT steers a PoC-generation agent with rule-based static analysis and with
dynamic feedback from sanitizers and coverage~\cite{desai_program_2026}.
PoC-Gym accepts a Java PoC only if its execution arrives at the vulnerable sink,
and finds that success markers alone are unreliable evidence that the
vulnerability fired~\cite{gezgin_poc-gym_2026}.
CyberGym~\cite{wang_cybergym_2025} and SEC-bench~\cite{lee_sec-bench_2025}
benchmark agents on reproducing known vulnerabilities, and the best agents
succeed on only about 20\% of tasks.
These systems start from a known vulnerability or code location.
\SystemName must first discover the bug, and its PoCs are tests that
run against each project's own mocks or testbenches.
Bug-injection techniques such as LAVA~\cite{dolan-gavitt_lava_2016},
FixReverter~\cite{zhang_fixreverter_2022}, and AVIATOR~\cite{aviator_2026},
and ground-truth benchmarks such as Magma~\cite{hazimeh_magma_2020}, could
support a future RoT benchmark (\S\ref{sec:disc-future}).

\parhead{LLM firmware analysis.}
On the firmware side, LLMs have been used to find vulnerabilities in IoT
devices.
FirmAgent uses fuzzing to find taint sources in IoT firmware, and LLM agents to
analyze the resulting paths and refine PoC test cases~\cite{ji_firmagent_2026}.
LIVA applies multi-agent LLM taint analysis to IoT device
binaries~\cite{yang_liva_2026}, and FirmHive coordinates a runtime tree of
agents over firmware images~\cite{zhang_llms_2025}.
These systems analyze IoT firmware binaries for input-handling bugs.
\SystemName instead analyzes RoT firmware source, where security-relevant bugs
often depend on boot-stage state and trust boundaries.

\parhead{Hardware security.}
LLMs are also increasingly applied to hardware security.
MARVEL uses a supervisor agent and tool-using executor agents to find RTL
vulnerabilities in a buggy OpenTitan-based SoC from the Hack@DATE
competition~\cite{marvel_2026}.
BugWhisperer fine-tunes an LLM to detect SoC
vulnerabilities~\cite{tarek_bugwhisperer_2025}, and SV-LLM coordinates agents
for SoC security verification, from threat modeling to simulation-based bug
validation~\cite{sv-llm_2025}.
BugGen synthesizes realistic RTL bugs~\cite{buggen_2025}.
HWE-Bench benchmarks agents on repairing real hardware bugs, including in
security roots of trust, and validates repairs through each project's own
simulation flows~\cite{hwe-bench_2026}.
HardSecBench measures the security awareness of LLMs that generate hardware
code~\cite{chen_hardsecbench_2026}.
Competition benchmarks such as Hack@DAC contain injected
bugs~\cite{dessouky_hardfails_2019}, and AttackonCTF shows that LLMs can
detect them through diff-style comparison rather than security
reasoning~\cite{rostami_attackonctf_2026}.
RTL fuzzers find processor bugs through differential
testing~\cite{hur_difuzzrtl_2021}, golden reference
models~\cite{kande_thehuzz_2022}, or formal
assistance~\cite{chen_hypfuzz_2023}.

To our knowledge, \SystemName is the first agentic system to ground
vulnerability discovery and review in project-specific RoT security
requirements and reproduce findings through native firmware and RTL test
infrastructure, uncovering previously unknown security-relevant bugs in
production RoT projects.

\section{Conclusion}
\label{sec:conclusion}

We presented \SystemName, a multi-stage agentic pipeline for discovering,
reviewing, and reproducing security-relevant bugs in RoT firmware and
SystemVerilog RTL.
By combining off-the-shelf models with a \texttt{Project Expert} agent,
per-project threat models, skeptical adversarial review, and toolchain-native
PoC generation, \SystemName filters candidate alerts and attaches executable
reproducers in each project's own build environment.
Across our evaluation scans on Caliptra, OpenTitan, and OpenPRoT, maintainers
have confirmed \evalConfirmedTotal of \SystemName's findings as previously
unknown security-relevant bugs (\evalConfirmedRt across three models on
Caliptra's runtime firmware and \evalConfirmedOther across single scans of the
other five targets).
On Caliptra's runtime firmware, it achieved
\evalRtArgonPrecision--\evalRtPrecision precision across three models
(\evalRtPrecision on Gemini~3.8 Flash, vs.\
\evalRtWholeDirNoTmPrecision--\evalRtWholeDirPrecision for the baselines).
On known bugs, it reached \evalKbFlashAllPct and \evalKbArgonAllPct recall on
Gemini~3.8 Flash and Gemini~4 Argon, compared to \evalKbFlashBaseAllPct and
\evalKbArgonBaseAllPct for the best baseline.

\SystemName demonstrates that grounding LLM agents in project-specific security
requirements and native test infrastructure enables the discovery, review, and
reproduction of security-relevant bugs in RoT firmware and RTL.
Executing PoCs on FPGAs, extending \SystemName to other RoT ecosystems, and
building contamination-free synthetic RoT benchmarks are natural next steps
toward continuous, evidence-backed security review of the firmware and hardware
that modern systems trust.

\section*{Acknowledgments}
\label{sec:acknowledgments}

We thank the Caliptra, OpenTitan, and OpenPRoT maintainers for triaging our
reports and confirming the issues that \SystemName found.
We also thank our colleagues at Google for their helpful feedback on earlier
drafts of this paper.

\section*{Ethical Considerations}
\label{sec:ethics}

\SystemName focuses on the discovery, review, and PoC reproduction of
security-relevant bugs in open-source RoT projects.
All potential and confirmed bugs discussed in this work were
responsibly disclosed to the maintainers of the affected projects prior to the
submission of this paper.
Some RoT components, such as boot ROM and RTL, cannot be fixed in deployed
silicon, so a disclosed bug may remain present in deployed devices.
We therefore withhold identifying details of new findings across targets in
this paper while upstream remediation and coordinated disclosure are in
progress.

\SystemName's pipeline includes the creation of PoC exploit code to reproduce
potential bugs.
The pipeline itself relies strictly on publicly available models and does not
introduce a new model or fine-tune or in any way change the weights of existing
models.
\SystemName also supports open-weight models run locally through Ollama, so
vulnerability analysis can be run without data being sent to a third-party
provider, if desired.
We only run PoC exploits on non-production machines, and never run them on
hardware or production deployments.
While the prompts and tooling that we use to steer models towards
PoC exploit creation could be abused by malicious actors, we
believe the benefits of vulnerability discovery outweigh the risk of abuse.
Attackers can already use general-purpose LLMs to produce exploits with
sufficient prompt engineering, and general-purpose vulnerability discovery
tools such as OpenAnt~\cite{korda_openant_2026} are already publicly
available.
In contrast, \SystemName enables CI-integrated reviews that help maintainers
catch security-relevant bugs before they reach released code.

Automated vulnerability scanners risk adding to the triage burden of
open-source maintainers if they flood repositories with unverified
alerts~\cite{baltes_endless_2026}.
\SystemName's skeptical review and toolchain-native PoC stages filter out
speculative alerts and attach executable reproducers before humans inspect a
report.
In addition, projects run \SystemName inside their own CI pipelines so findings
arrive through their normal code-review workflow rather than as unsolicited
external submissions.

\bibliographystyle{plainurl}
\bibliography{conferences,refs,refs-paper}

\appendix
\section{Agent Tools}
\label{sec:appendix-tools}

Table~\ref{tab:tools} describes each tool available to \SystemName's agents,
grouped by the categories used in Table~\ref{tab:agent-tools}.

\begin{table}[!ht]
    \centering
    \footnotesize
    \newcommand{\cat}[1]{\multicolumn{2}{@{}l}{\textbf{#1}} \\}
    \begin{tabularx}{\columnwidth}{@{}l >{\raggedright\arraybackslash}X@{}}
        \toprule
        \cat{File access}
        \texttt{glob}                 & Lists file paths that match a glob pattern. \\
        \texttt{read\_file}           & Reads a file, optionally limited to a line range. \\
        \midrule
        \cat{Code search}
        \texttt{ctags\_search}        & Looks up symbol definitions in a Universal Ctags index. \\
        \texttt{ast\_search}          & Finds structural code patterns with ast-grep. \\
        \texttt{grep\_search}         & Searches file contents for a pattern with ripgrep. \\
        \midrule
        \cat{CWE lookup}
        \texttt{search\_cwe}          & Searches the bundled MITRE CWE catalog by keyword. \\
        \midrule
        \cat{Expert query}
        \texttt{ask\_project\_expert} & Asks the \texttt{Project Expert} agent a project-level question. \\
        \midrule
        \cat{Worktree}
        \texttt{write\_worktree\_file} & Creates or overwrites a file in the scratch worktree. \\
        \texttt{patch\_worktree\_file} & Replaces an exact block of text in a worktree file. \\
        \texttt{run\_harness\_command} & Runs a build or test command (\harnessTimeout timeout). \\
        \texttt{get\_worktree\_diff}  & Returns the diff of all changes in the worktree. \\
        \texttt{reset\_worktree}      & Restores the worktree to its clean baseline. \\
        \bottomrule
    \end{tabularx}
    \caption[Agent tools by category.]{Agent tools by category. Tool output is truncated to \toolOutputMax
    characters.}
    \label{tab:tools}
\end{table}

\section{Cost Analysis}
\label{sec:appendix-cost}

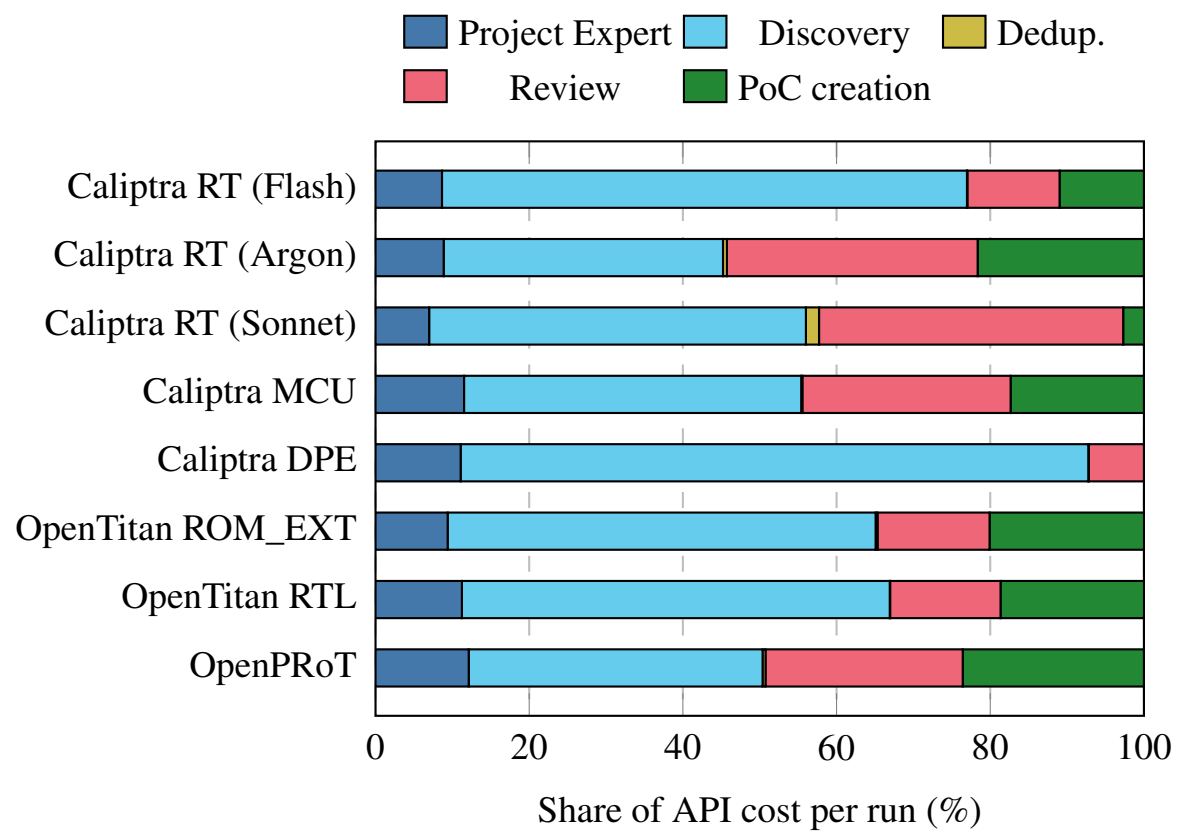
\begin{figure}[!ht]
    \centering
    \begin{tikzpicture}
\begin{axis}[
    xbar stacked, bar width=7pt, y dir=reverse,
    scale only axis, width=0.6\columnwidth, height=3.8cm,
    xmin=0, xmax=100, xmajorgrids=true, grid style=dashed,
    xlabel={Share of API cost per run (\%)},
    ytick={0,...,7}, yticklabels={{Caliptra RT (Flash)},{Caliptra RT (Argon)},{Caliptra RT (Sonnet)},{Caliptra MCU},{Caliptra DPE},{OpenTitan ROM\_EXT},{OpenTitan RTL},{OpenPRoT}},
    yticklabel style={font=\scriptsize}, xticklabel style={font=\scriptsize},
    xlabel style={font=\scriptsize},
    legend style={at={(0.5,1.02)}, anchor=south, legend columns=3, font=\scriptsize, draw=none},
]
\addplot[fill={rgb,255:red,68;green,119;blue,170}, draw=black] coordinates { (8.65,0) (8.87,1) (6.98,2) (11.52,3) (11.08,4) (9.39,5) (11.24,6) (12.13,7) };
\addlegendentry{Project Expert}
\addplot[fill={rgb,255:red,102;green,204;blue,238}, draw=black] coordinates { (68.32,0) (36.35,1) (49.03,2) (43.88,3) (81.68,4) (55.69,5) (55.68,6) (38.24,7) };
\addlegendentry{Discovery}
\addplot[fill={rgb,255:red,204;green,187;blue,68}, draw=black] coordinates { (0.05,0) (0.51,1) (1.71,2) (0.15,3) (0.04,4) (0.25,5) (0.05,6) (0.38,7) };
\addlegendentry{Dedup.}
\addplot[fill={rgb,255:red,238;green,102;blue,119}, draw=black] coordinates { (12.00,0) (32.63,1) (39.58,2) (27.10,3) (7.20,4) (14.58,5) (14.37,6) (25.67,7) };
\addlegendentry{Review}
\addplot[fill={rgb,255:red,34;green,136;blue,51}, draw=black] coordinates { (10.98,0) (21.63,1) (2.71,2) (17.35,3) (0.00,4) (20.10,5) (18.67,6) (23.58,7) };
\addlegendentry{PoC creation}
\end{axis}
\end{tikzpicture}
    \caption{Share of API cost per run by pipeline stage.}
    \label{fig:phase-cost}
\end{figure}

This appendix breaks down the cost of \SystemName's runs on Caliptra's runtime
firmware (\S\ref{sec:eval-cost}) by token type, pipeline stage, and model.
Table~\ref{tab:rt-tokens} reports, per run and averaged over three runs, the
input tokens, cache hit rate, tool calls, thinking and output tokens,
list-price API cost, and wall-clock time of each configuration.
Figure~\ref{fig:phase-cost} shows how each model's cost splits across pipeline
stages on Caliptra's runtime firmware (RT), and how Gemini~3.8 Flash's cost
splits on the other targets.

\begin{table*}[!ht]
    \footnotesize
    \centering
    \setlength{\tabcolsep}{3.8pt}
    \caption{Mean token use, tool calls, cost, and time per run on Caliptra's
    runtime firmware.}
    \label{tab:rt-tokens}
    \begin{tabular}{@{}lrrrrrrr@{}}
        \toprule
        \textbf{Configuration} & \textbf{Input} & \textbf{Cache} & \textbf{Tools} & \textbf{Thinking} & \textbf{Output} & \textbf{Cost} & \textbf{Time} \\
        \midrule
        \SystemName & \evalRtTotalInput & \evalRtCachePct & \evalRtToolCalls & \evalRtThinking & \evalRtOutput & \evalRtCost & \evalRtTime \\
        \texttt{w/o Threat Model} & \evalRtNoTmTotalInput & \evalRtNoTmCachePct & \evalRtNoTmToolCalls & \evalRtNoTmThinking & \evalRtNoTmOutput & \evalRtNoTmCost & \evalRtNoTmTime \\
        \texttt{w/o Project Expert} & \evalRtNoExpertTotalInput & \evalRtNoExpertCachePct & \evalRtNoExpertToolCalls & \evalRtNoExpertThinking & \evalRtNoExpertOutput & \evalRtNoExpertCost & \evalRtNoExpertTime \\
        Single-Agent (Per-File, w/o TM) & \evalRtPerFileNoTmTotalInput & \evalRtPerFileNoTmCachePct & \evalRtPerFileNoTmToolCalls & \evalRtPerFileNoTmThinking & \evalRtPerFileNoTmOutput & \evalRtPerFileNoTmCost & \evalRtPerFileNoTmTime \\
        Single-Agent (Per-File) & \evalRtPerFileTotalInput & \evalRtPerFileCachePct & \evalRtPerFileToolCalls & \evalRtPerFileThinking & \evalRtPerFileOutput & \evalRtPerFileCost & \evalRtPerFileTime \\
        Single-Agent (Whole-Dir, w/o TM) & \evalRtWholeDirNoTmTotalInput & \evalRtWholeDirNoTmCachePct & \evalRtWholeDirNoTmToolCalls & \evalRtWholeDirNoTmThinking & \evalRtWholeDirNoTmOutput & \evalRtWholeDirNoTmCost & \evalRtWholeDirNoTmTime \\
        Single-Agent (Whole-Dir) & \evalRtWholeDirTotalInput & \evalRtWholeDirCachePct & \evalRtWholeDirToolCalls & \evalRtWholeDirThinking & \evalRtWholeDirOutput & \evalRtWholeDirCost & \evalRtWholeDirTime \\
        \SystemName (Gemini 4 Argon) & \evalRtArgonTotalInput & \evalRtArgonCachePct & \evalRtArgonToolCalls & \evalRtArgonThinking & \evalRtArgonOutput & \evalRtArgonCost & \evalRtArgonTime \\
        \SystemName (Claude Sonnet 5.5) & \evalRtSonnetTotalInput & \evalRtSonnetCachePct & \evalRtSonnetToolCalls & \evalRtSonnetThinking & \evalRtSonnetOutput & \evalRtSonnetCost & \evalRtSonnetTime \\
        \bottomrule
    \end{tabular}
\end{table*}

On Caliptra's runtime firmware, uncached input tokens dominate API cost across
configurations (Table~\ref{tab:rt-tokens}), so a model's tool-call volume and
cache hit rate matter more than its per-token list price.
Although Gemini~4 Argon has higher list prices than Gemini~3.8 Flash
(\priceArgonUncachedIn vs.\ \priceFlashUncachedIn per million uncached input
tokens), Flash makes more LLM calls (\evalRtLlmCalls vs.\
\evalRtArgonLlmCalls) and tool calls (\evalRtFileToolCalls file reads and
\evalRtSearchToolCalls code searches), reading \evalRtFlashArgonInputRatio as
many input tokens per run with a lower cache hit rate (\evalRtCachePct vs.\
\evalRtArgonCachePct) while reporting fewer open findings (\evalRtOpenMean
vs.\ \evalRtArgonOpenMean).
Claude Sonnet~5.5 makes the fewest tool calls (\evalRtSonnetToolCalls per run),
reading the fewest tokens and finishing fastest (\evalRtSonnetTime) at the
lowest cost (\evalRtSonnetCost).
Removing the \texttt{Project Expert} agent increases cost from \evalRtCost to
\evalRtNoExpertCost per run because the \texttt{Source File Auditor} and
\texttt{Adversarial Reviewer} agents must explore the repository themselves in
separate sessions with little cache reuse (\evalRtNoExpertAuditorToolCalls and
\evalRtNoExpertReviewerToolCalls tool calls per run vs.\
\evalRtAuditorToolCalls and \evalRtReviewerToolCalls with it).

Discovery accounts for the largest share of spend across every model and target
(Figure~\ref{fig:phase-cost}), representing \evalRtAuditorCostPct of Flash's
cost per run (\evalRtAuditorCost of \evalRtCost) and \evalRtArgonAuditorCostPct
of Argon's.
Flash also spends most of its output on thinking, with \evalRtThinking
thinking tokens and \evalRtOutput output tokens per run
(Table~\ref{tab:rt-tokens}).

\section{Model Safeguard Refusals}
\label{sec:appendix-safeguards}

\SystemName logs four categories of safeguard and output-validation events
during execution (Table~\ref{tab:safeguard-refusals}): provider-level safety
filter blocks returned by the model API (\emph{API}), plain-text safety-policy
refusals emitted in place of structured output (\emph{Hard}), valid structured
responses that populate an optional \texttt{refusal\_reason} field rather than
completing the task (\emph{Soft}), and non-refusal outputs rejected by Pydantic
schema validation (\emph{Drops}).
\emph{Total} sums the three kinds of safety refusal, and the stability row
lists only the two runs beyond the first three so each run is counted once.
Refusals are rare overall (\evalTotalSafetyRefusals across the
\evalTotalUniqueRuns runs on our six evaluation targets) and concentrate in the
\texttt{PoC Exploit Generator} and \texttt{Adversarial Reviewer} agents during
full-pipeline runs, and in the \texttt{Source File Auditor} agent when the
\texttt{Project Expert} agent is ablated.
All \evalTotalSchemaValidationDrops schema-validation drops occurred in the
\texttt{Source File Auditor} agent under Claude Sonnet~5.5.

\begin{table}[!ht]
    \centering
    \footnotesize
    \setlength{\tabcolsep}{2.2pt}
    \begin{tabular*}{\columnwidth}{@{}l@{\extracolsep{\fill}}rrrrrr@{}}
        \toprule
        \textbf{Target / Configuration} & \textbf{Runs} & \textbf{API} & \textbf{Hard} & \textbf{Soft} & \textbf{Total} & \textbf{Drops} \\
        \midrule
        \multicolumn{7}{@{}l@{}}{\textit{Caliptra runtime firmware}} \\
        \hspace{0.5em}\SystemName & 3 & \evalRtRefusalsApi & \evalRtRefusalsHard & \evalRtRefusalsSoft & \evalRtRefusalsTotal & \evalRtSchemaDrops \\
        \hspace{0.5em}\SystemName (+2 stability) & 2 & \evalRtStabilityExtraRefusalsApi & \evalRtStabilityExtraRefusalsHard & \evalRtStabilityExtraRefusalsSoft & \evalRtStabilityExtraRefusalsTotal & \evalRtStabilityExtraSchemaDrops \\
        \hspace{0.5em}\texttt{w/o Threat Model} & 3 & \evalRtNoTmRefusalsApi & \evalRtNoTmRefusalsHard & \evalRtNoTmRefusalsSoft & \evalRtNoTmRefusalsTotal & \evalRtNoTmSchemaDrops \\
        \hspace{0.5em}\texttt{w/o Project Expert} & 3 & \evalRtNoExpertRefusalsApi & \evalRtNoExpertRefusalsHard & \evalRtNoExpertRefusalsSoft & \evalRtNoExpertRefusalsTotal & \evalRtNoExpertSchemaDrops \\
        \hspace{0.5em}Per-File (w/o TM) & 3 & \evalRtPerFileNoTmRefusalsApi & \evalRtPerFileNoTmRefusalsHard & \evalRtPerFileNoTmRefusalsSoft & \evalRtPerFileNoTmRefusalsTotal & \evalRtPerFileNoTmSchemaDrops \\
        \hspace{0.5em}Per-File & 3 & \evalRtPerFileRefusalsApi & \evalRtPerFileRefusalsHard & \evalRtPerFileRefusalsSoft & \evalRtPerFileRefusalsTotal & \evalRtPerFileSchemaDrops \\
        \hspace{0.5em}Whole-Dir (w/o TM) & 3 & \evalRtWholeDirNoTmRefusalsApi & \evalRtWholeDirNoTmRefusalsHard & \evalRtWholeDirNoTmRefusalsSoft & \evalRtWholeDirNoTmRefusalsTotal & \evalRtWholeDirNoTmSchemaDrops \\
        \hspace{0.5em}Whole-Dir & 3 & \evalRtWholeDirRefusalsApi & \evalRtWholeDirRefusalsHard & \evalRtWholeDirRefusalsSoft & \evalRtWholeDirRefusalsTotal & \evalRtWholeDirSchemaDrops \\
        \hspace{0.5em}\SystemName (Gemini 4 Argon) & 3 & \evalRtArgonRefusalsApi & \evalRtArgonRefusalsHard & \evalRtArgonRefusalsSoft & \evalRtArgonRefusalsTotal & \evalRtArgonSchemaDrops \\
        \hspace{0.5em}\SystemName (Claude Sonnet 5.5) & 3 & \evalRtSonnetRefusalsApi & \evalRtSonnetRefusalsHard & \evalRtSonnetRefusalsSoft & \evalRtSonnetRefusalsTotal & \evalRtSonnetSchemaDrops \\
        \midrule
        \multicolumn{7}{@{}l@{}}{\textit{Other targets (Gemini 3.8 Flash)}} \\
        \hspace{0.5em}Caliptra MCU ROM, protocols & 1 & \evalCaliptraMcuSwRefusalsApi & \evalCaliptraMcuSwRefusalsHard & \evalCaliptraMcuSwRefusalsSoft & \evalCaliptraMcuSwRefusalsTotal & \evalCaliptraMcuSwSchemaDrops \\
        \hspace{0.5em}Caliptra DPE & 1 & \evalCaliptraDpeRefusalsApi & \evalCaliptraDpeRefusalsHard & \evalCaliptraDpeRefusalsSoft & \evalCaliptraDpeRefusalsTotal & \evalCaliptraDpeSchemaDrops \\
        \hspace{0.5em}OpenTitan ROM\_EXT firmware & 1 & \evalOpentitanFwRefusalsApi & \evalOpentitanFwRefusalsHard & \evalOpentitanFwRefusalsSoft & \evalOpentitanFwRefusalsTotal & \evalOpentitanFwSchemaDrops \\
        \hspace{0.5em}OpenTitan Earl Grey RTL & 1 & \evalOpentitanRtlRefusalsApi & \evalOpentitanRtlRefusalsHard & \evalOpentitanRtlRefusalsSoft & \evalOpentitanRtlRefusalsTotal & \evalOpentitanRtlSchemaDrops \\
        \hspace{0.5em}OpenPRoT platform services & 1 & \evalOpenprotRefusalsApi & \evalOpenprotRefusalsHard & \evalOpenprotRefusalsSoft & \evalOpenprotRefusalsTotal & \evalOpenprotSchemaDrops \\
        \midrule
        \textbf{Total} & \evalTotalUniqueRuns & \evalTotalApiSafetyBlocks & \evalTotalHardSafetyRefusals & \evalTotalSoftRefusals & \evalTotalSafetyRefusals & \evalTotalSchemaValidationDrops \\
        \bottomrule
    \end{tabular*}
    \caption{Model safeguard refusals and schema-validation drops per
    configuration and target, summed across runs.}
    \label{tab:safeguard-refusals}
\end{table}

\section{Known Caliptra Bugs}
\label{sec:appendix-known-bugs}

Table~\ref{tab:known-bugs-list} lists the 34 known security-relevant bugs in
Caliptra's core firmware at \texttt{fw-2.1.0} against which we measure recall
in RQ4 (\S\ref{sec:eval-known-bugs}).
Each bug was fixed upstream in April 2026 or later (across 33 pull requests,
as \texttt{\#4148} fixes both \texttt{SB-18} and \texttt{SH-05}), after both
Gemini models' training cutoff.
In Table~\ref{tab:known-bugs-list}, \emph{Sev.} is the upstream advisory
severity (Tier~A only; Tiers~B and~C carry no upstream advisory), \emph{CWE}
omits the \texttt{CWE-} prefix (assigned by us across all three tiers under
\S\ref{sec:threat-model}, as most Tier~A GitHub advisories list no CWE),
\emph{Fix} is the upstream pull request, and \emph{Location} lists the affected
files at \texttt{fw-2.1.0}.

\begin{table*}[t]
    \scriptsize
    \centering
    \setlength{\tabcolsep}{3pt}
    \caption{Known security-relevant bugs in Caliptra's core firmware at \texttt{fw-2.1.0} (commit \texttt{7da3b01}).}
    \label{tab:known-bugs-list}
    \begin{tabular}{@{}lllr>{\raggedright\arraybackslash}p{0.32\textwidth}>{\raggedright\arraybackslash}p{0.40\textwidth}@{}}
        \toprule
        ID & Sev. & CWE & Fix & Location & Description \\
        \midrule
        \multicolumn{6}{@{}l}{\textit{Tier A: published CVEs (5)}} \\
        CVE-2026-5818 & High & 287 & \#3719 & \texttt{runtime/\allowbreak{}src/\allowbreak{}activate\_firmware.rs} & MCU Firmware Update Authentication Bypass on Caliptra Core \\
        CVE-2026-6458 & Medium & 347 & \#3790 & \texttt{drivers/\allowbreak{}src/\allowbreak{}aes.rs} & AES-256-GCM Authentication Tag Does Not Cover First Ciphertext Blocks When AAD Is Empty \\
        CVE-2026-7328 & Medium & 862 & \#3864 & \texttt{runtime/\allowbreak{}src/\allowbreak{}cryptographic\_mailbox.rs} \newline \texttt{runtime/\allowbreak{}src/\allowbreak{}invoke\_dpe.rs} \newline \texttt{runtime/\allowbreak{}src/\allowbreak{}lib.rs} & Unverified AXI Address / Missing PL0 Check in Subsystem Mode Commands Enables DoS / Arbitrary DMA \\
        CVE-2026-11835 & Medium & 367 & \#3702 & \texttt{rom/\allowbreak{}dev/\allowbreak{}src/\allowbreak{}flow/\allowbreak{}cold\_reset/\allowbreak{}fw\_processor/\allowbreak{}mod.rs} \newline \texttt{rom/\allowbreak{}dev/\allowbreak{}src/\allowbreak{}flow/\allowbreak{}update\_reset.rs} & Caliptra Update-Reset / Subsystem Secure-Boot Bypass via External AXI Staging TOCTOU \\
        CVE-2026-11836 & Low & 290 & \#3694 & \texttt{common/\allowbreak{}src/\allowbreak{}debug\_unlock.rs} & Production Debug-Unlock Token Verification Missing Device Binding \\
        \midrule
        \multicolumn{6}{@{}l}{\textit{Tier B: security bugs fixed by maintainers without a CVE (22)}} \\
        SB-01 & -- & 287 & \#3602 & \texttt{common/\allowbreak{}src/\allowbreak{}debug\_unlock.rs} & Production Debug Unlock Accepted When Public-Key Hash Fuse Is Uninitialized or Zeroized \\
        SB-02 & -- & 326 & \#3606 & \texttt{runtime/\allowbreak{}src/\allowbreak{}ocp\_lock/\allowbreak{}initialize\_mek\_secret.rs} \newline \texttt{runtime/\allowbreak{}src/\allowbreak{}ocp\_lock/\allowbreak{}mod.rs} & OCP-LOCK Accepts All-Zero (Zeroized) SEK and DPK Secrets \\
        SB-03 & -- & 840 & \#3619 & \texttt{runtime/\allowbreak{}src/\allowbreak{}activate\_firmware.rs} & ACTIVATE\_FIRMWARE Indexes Activation Bitmap by fw\_id Instead of exec\_bit \\
        SB-04 & -- & 841 & \#3628 & \texttt{runtime/\allowbreak{}src/\allowbreak{}debug\_unlock.rs} & Runtime Production Debug Unlock Challenge Omits PROD\_DBG\_UNLOCK\_IN\_PROGRESS Hardware Bit \\
        SB-05 & -- & 670 & \#3675 & \texttt{common/\allowbreak{}src/\allowbreak{}wdt.rs} & stop\_wdt Fails to Stop Watchdog Timer After Production Debug Unlock \\
        SB-06 & -- & 345 & \#3688 & \texttt{runtime/\allowbreak{}src/\allowbreak{}authorize\_and\_stash.rs} & AUTHORIZE\_AND\_STASH Stashes Unchecked Caller Measurement Instead of Verified DMA Digest \\
        SB-07 & -- & 125 & \#3732 & \texttt{runtime/\allowbreak{}src/\allowbreak{}manifest.rs} & Auth Manifest Binary Search Ignores entry\_count Over Fixed Backing Array \\
        SB-08 & -- & 862 & \#3734 & \texttt{runtime/\allowbreak{}src/\allowbreak{}set\_auth\_manifest.rs} & SET\_AUTH\_MANIFEST Missing PL0 PAUSER Privilege Check \\
        SB-09 & -- & 367 & \#3766 & \texttt{common/\allowbreak{}src/\allowbreak{}debug\_unlock.rs} & Production Debug Unlock Public-Key Hash Computed Directly Over Shared Mailbox SRAM (TOCTOU) \\
        SB-10 & -- & 284 & \#3804 & \texttt{rom/\allowbreak{}dev/\allowbreak{}src/\allowbreak{}flow/\allowbreak{}uds\_programming.rs} & ROM UDS Programming Permitted While Debug Intent Strap Is Asserted \\
        SB-11 & -- & 347 & \#3803 & \texttt{rom/\allowbreak{}dev/\allowbreak{}src/\allowbreak{}flow/\allowbreak{}cold\_reset/\allowbreak{}fw\_processor/\allowbreak{}ecdsa\_verify.rs} \newline \texttt{runtime/\allowbreak{}src/\allowbreak{}lib.rs} & Hash-Based ECDSA384 and LMS Signature Verification Improperly Report FIPS\_APPROVED \\
        SB-12 & -- & 252 & \#3858 & \texttt{common/\allowbreak{}src/\allowbreak{}uds\_fe\_programming.rs} & Silent Discard of OTP Verification Errors During UDS/FE Zeroization \\
        SB-13 & -- & 276 & \#3873 & \texttt{rom/\allowbreak{}dev/\allowbreak{}src/\allowbreak{}flow/\allowbreak{}cold\_reset/\allowbreak{}ldev\_id.rs} \newline \texttt{rom/\allowbreak{}dev/\allowbreak{}src/\allowbreak{}flow/\allowbreak{}update\_reset.rs} \newline \texttt{rom/\allowbreak{}dev/\allowbreak{}src/\allowbreak{}flow/\allowbreak{}warm\_reset.rs} & ROM Fails to Write-Lock Stable IDevID and LDevID Identity Root Keys in Key Vault \\
        SB-14 & -- & 862 & \#4131 & \texttt{runtime/\allowbreak{}src/\allowbreak{}activate\_firmware.rs} \newline \texttt{runtime/\allowbreak{}src/\allowbreak{}fe\_programming.rs} & FE\_PROG and ACTIVATE\_FIRMWARE Missing PL0 PAUSER Privilege Checks \\
        SB-15 & -- & 320 & \#4111 & \texttt{runtime/\allowbreak{}src/\allowbreak{}cryptographic\_mailbox.rs} & CM\_AES\_ENCRYPT\_INIT and CM\_AES\_DECRYPT\_INIT Omit CMK key\_usage Check \\
        SB-16 & -- & 326 & \#4141 & \texttt{runtime/\allowbreak{}src/\allowbreak{}ocp\_lock/\allowbreak{}mod.rs} & OCP-LOCK AES-XTS MEK Generation and Unwrap Omit Key-Half Distinctness Check \\
        SB-17 & -- & 862 & \#4197 & \texttt{runtime/\allowbreak{}src/\allowbreak{}pcr.rs} & EXTEND\_PCR Missing PL0 PAUSER Privilege Check and PCR31 Reservation \\
        SB-18 & -- & 347 & \#4148 & \texttt{common/\allowbreak{}src/\allowbreak{}debug\_unlock.rs} \newline \texttt{common/\allowbreak{}src/\allowbreak{}verify.rs} \newline \texttt{runtime/\allowbreak{}src/\allowbreak{}cryptographic\_mailbox.rs} & Missing Trailing Padding Byte Validation in ML-DSA-87 Signature Verification \\
        SB-19 & -- & 862 & \#4252 & \texttt{runtime/\allowbreak{}src/\allowbreak{}cryptographic\_mailbox.rs} & CM\_DERIVE\_STABLE\_KEY Missing PL0 PAUSER Privilege Check \\
        SB-20 & -- & 345 & \#4215 & \texttt{image/\allowbreak{}verify/\allowbreak{}src/\allowbreak{}verifier.rs} & ROM ImageVerifier Missing FMC Offset Contiguity and Update-Reset FMC Load-Address/Size Checks \\
        SB-21 & -- & 822 & \#4224 & \texttt{image/\allowbreak{}verify/\allowbreak{}src/\allowbreak{}verifier.rs} & ROM ImageVerifier Validates FMC and Runtime Entry Points Against Global ICCM Rather Than Image Bounds \\
        SB-22 & -- & 367 & \#4271 & \texttt{common/\allowbreak{}src/\allowbreak{}verifier.rs} \newline \texttt{drivers/\allowbreak{}src/\allowbreak{}dma.rs} \newline \texttt{image/\allowbreak{}verify/\allowbreak{}src/\allowbreak{}verifier.rs} & ROM Firmware Manifest Snapshot TOCTOU, AXI 64-Bit Address Overflow, Runtime Gap/Bundle Bounds, and SVN Cap Bypass \\
        \midrule
        \multicolumn{6}{@{}l}{\textit{Tier C: zeroization, attestation, and side-channel hardening fixes (7)}} \\
        SH-01 & -- & 226 & \#3807 & \texttt{rom/\allowbreak{}dev/\allowbreak{}src/\allowbreak{}flow/\allowbreak{}cold\_reset/\allowbreak{}fw\_processor/\allowbreak{}mod.rs} \newline \texttt{rom/\allowbreak{}dev/\allowbreak{}src/\allowbreak{}flow/\allowbreak{}cold\_reset/\allowbreak{}fw\_processor/\allowbreak{}shutdown.rs} & ROM SHUTDOWN Command Acknowledges Mailbox Response Before Zeroizing Critical Security Parameters \\
        SH-02 & -- & 226 & \#3808 & \texttt{runtime/\allowbreak{}src/\allowbreak{}fips.rs} & Runtime FIPS Shutdown Omits ML-KEM/ML-DSA Zeroization and Ignores TRNG Zeroize Errors \\
        SH-03 & -- & 290 & \#3930 & \texttt{common/\allowbreak{}src/\allowbreak{}dice.rs} \newline \texttt{rom/\allowbreak{}dev/\allowbreak{}src/\allowbreak{}pcr.rs} & ROM DeviceStatus PCR0 and FMC Alias DICE Measurements Omit Subsystem Operating Mode \\
        SH-04 & -- & 330 & \#4106 & \texttt{rom/\allowbreak{}dev/\allowbreak{}src/\allowbreak{}main.rs} & ROM Consumes CSRNG Output for CFI Initialization Before Running CsrngKat Self-Test \\
        SH-05 & -- & 226 & \#4148 & \texttt{common/\allowbreak{}src/\allowbreak{}crypto.rs} \newline \texttt{drivers/\allowbreak{}src/\allowbreak{}hpke/\allowbreak{}kem/\allowbreak{}mlkem.rs} \newline \texttt{drivers/\allowbreak{}src/\allowbreak{}ml\_kem.rs} & Missing Stack Zeroization of Decrypted CMK and ML-KEM Secret Key Material \\
        SH-06 & -- & 226 & \#4240 & \texttt{drivers/\allowbreak{}src/\allowbreak{}hpke/\allowbreak{}kem/\allowbreak{}p384.rs} & Missing Stack Zeroization of Raw P-384 ECDH Shared Secret in HPKE KEM \\
        SH-07 & -- & 755 & \#4304 & \texttt{fmc/\allowbreak{}src/\allowbreak{}main.rs} & FMC Invalid Handoff Table Fallback Calls ExitCtrl::exit Instead of handle\_fatal\_error \\
        \bottomrule
    \end{tabular}
\end{table*}

\end{document}